\documentclass[lettersize,journal]{IEEEtran}
\usepackage{amsthm,amsmath,amssymb}
\usepackage{caption}
\usepackage{graphicx}
\usepackage{booktabs}
\usepackage{tabularray}
\usepackage{stfloats}
\usepackage[caption=false,font=footnotesize]{subfig}
\usepackage[font=small]{caption}
\usepackage{cite}
\usepackage{booktabs}
\usepackage{soul}
\usepackage{textcomp}
\usepackage{mathtools}
\usepackage{color}
\usepackage{algorithm}
\usepackage{algorithmic}
\usepackage{xcolor}

\usepackage{array}

\UseTblrLibrary{booktabs,siunitx}
\DeclareSIUnit{\belmilliwatt}{Bm}
\DeclareSIUnit{\dBm}{\deci\belmilliwatt}
\DeclareSIUnit{\rad}{rad}

\newtheorem{remark}{Remark}

\allowdisplaybreaks

\begin{document}
 
\title{Near-Field Multiuser Localization Based on Extremely Large Antenna Array with Limited RF Chains}
\author{Boyu~Teng,~\IEEEmembership{Student~Member,~IEEE},~Xiaojun~Yuan,~\IEEEmembership{Senior~Member,~IEEE},~Rui~Wang,~\IEEEmembership{Senior~Member,~IEEE},\\Ying-Chang~Liang,~\IEEEmembership{Fellow,~IEEE},~and~Xinming~Huang
	\thanks{B. Teng and X. Yuan are with the National Key Laboratory of Wireless Communications, University of Electronic Science and Technology of China, Chengdu 611731, China (e-mail: byteng@std.uestc.edu.cn; xjyuan@uestc.edu.cn).}
	\thanks{R. Wang is with the College of Electronics and Information Engineering, Tongji University, Shanghai 201804, China. R. Wang is also with the Shanghai Institute of Intelligent Science and Technology, Tongji University, Shanghai 201804, China (e-mail: ruiwang@tongji.edu.cn).
}
\thanks{Y.-C. Liang is with the Center for Intelligent Networking and Communications (CINC),
University of Electronic Science and Technology of China (UESTC), Chengdu
611731, China (e-mail: liangyc@ieee.org).
}
\thanks{X. Huang is with the College of Electronic Science and Technology, National University of Defense Technology, Changsha 410073, China, and also with the Key Laboratory of Satellite Navigation Technology, Changsha 410073, China (e-mail: hxm\_kd@163.com).}
\thanks{This work was supported in part by the National Key Laboratory of Wireless Communications Foundation under Grant IFN20230204, and in part by Hunan Hejian Innovative Talent project (Project number: 2024RC3121), and in part by independent research projects of the Key Laboratory of Satellite Navigation Technology (Project number: WXDHS2023105).}
}
\maketitle
\begin{abstract}

Extremely {large antenna array} (ELAA) not only effectively enhances system communication performance but also improves the sensing capabilities of communication systems, making it one of the key enabling technologies in 6G wireless networks. This paper investigates the multiuser localization problem in an uplink Multiple Input Multiple Output (MIMO) system, where the base station (BS) is equipped with an ELAA to receive signals from multiple single-antenna users. We exploit analog beamforming to reduce the number of radio frequency (RF) chains.
We first develop a comprehensive near-field ELAA channel model that accounts for the antenna radiation pattern and free space path loss. Due to the large aperture of the ELAA, the angular resolution of the array is high, which improves user localization accuracy. However, it also makes the user localization problem highly non-convex, posing significant challenges when the number of RF chains is limited. To address this issue, we use an array partitioning strategy to divide the ELAA channel into multiple subarray channels and utilize the geometric constraints between user locations and subarrays for probabilistic modeling.
To fully exploit these geometric constraints, we propose the array partitioning-based location estimation with limited measurements (APLE-LM) algorithm based on the message passing principle to achieve multiuser localization. We derive the Bayesian Cramér-Rao Bound (BCRB) as the theoretical performance lower bound for our formulated near-field multiuser localization problem. 
Extensive simulations under various parameter configurations validate the proposed APLE-LM algorithm. The results demonstrate that APLE-LM achieves superior localization accuracy compared to baseline algorithms and approaches the BCRB at high signal-to-noise ratio (SNR).
\end{abstract}
\begin{IEEEkeywords}
User localization,
  large antenna array,
  near-field,
  limited RF chains,
  Bayesian inference.
\end{IEEEkeywords}
\section{Introduction}
The sixth-generation (6G) mobile communications system promises unprecedented advancements in connectivity, speed, and reliability \cite{jiang2021road,chowdhury20206g}. A critical aspect of realizing the full potential of 6G lies in the accurate localization of user equipments (UEs) \cite{de2021convergent}. High-precision UE localization is vital for numerous applications, including autonomous driving, augmented reality (AR), and advanced Internet of Things (IoT) systems\cite{de2021convergent,trevlakis2023localization}. The need for precise UE localization in 6G is further driven by the demand for ultra-reliable and low-latency communication (URLLC), enabling seamless and robust interactions between devices in complex environments. Recently, integrated sensing and communication (ISAC) has garnered significant attention \cite{wymeersch2021integration,liu2022integrated,liu2022survey}. ISAC leverages existing wireless communication infrastructure for radio localization, embedding localization as an intrinsic feature of the communication system without the need for additional dedicated sensor hardware\cite{wymeersch2021integration}. Traditional radio localization methods typically rely on multiple sites as anchors, using collaborative measurements such as angle of arrival (AoA) and time of arrival (ToA) to estimate UE positions \cite{teng2022bayesian,wymeersch2009cooperative,motie2024self}. However, real-world deployments of multi-site collaboration face significant challenges due to non-ideal factors such as synchronization errors and carrier frequency offsets.
\par
{As an advanced evolution of massive multiple-input multiple-output (mMIMO) technology, extremely large antenna arrays (ELAA) have attracted significant research attention due to their potential to enhance communication system performance and provide novel opportunities for radio localization\cite{wang2024tutorial,cui2022near,an2023toward,wang2023extremely,zhang2023near,lu2023near,cui2022channel}. Compared to conventional mMIMO, ELAA involves a substantial increase in the number of antennas (e.g., more than thousands) and introduces fundamental changes in electromagnetic (EM) characteristics. In mMIMO systems, the base station (BS) antenna array typically has a small aperture, and the UE is assumed to be located in the far-field region, where the curvature of the electromagnetic wavefront arriving at the array is negligible. This allows the wavefront to be approximated as a plane \cite{teng2022bayesian,wang2021joint}. However, as the number of antennas and the array aperture increase in ELAA, the UE can be located in the near-field region of the ELAA, where the curvature of the electromagnetic wavefront becomes significant \cite{cui2022near,wang2023extremely}. In this scenario, a near-field spherical wavefront model is required, instead of the far-field plane wavefront model used in mMIMO systems. In the spherical wavefront model, the received signal depends on the incident angle and the distance between the UE and ELAA.} This near-field effect enables precise distance estimation by leveraging the varying wavefront curvatures at different distances, even for signals arriving from the same direction.
Furthermore, with its larger array size and increased number of antenna elements, ELAA significantly enhances angular resolution and angle estimation accuracy, enabling high-precision single-site localization using both angle and distance information.
\par
To fully exploit the ELAA's high angular resolution and the spherical wave characteristics of electromagnetic propagation in localization, extensive research has been conducted on the near-field localization problem \cite{yuan2024scalable,teng2024near,zhang2018localization,zhi2007near,pan2023ris,qiao2023sensing,abu2021near,chen2024elaa,lu22023near}. Among them, the prior work \cite{yuan2024scalable} proposed the array partitioning-based location estimation (APLE) algorithm, where the ELAA is partitioned into multiple subarrays. The AoAs at each subarray are estimated independently using the far-field channel model. Then the UE position is estimated based on geometric constraints between the AoAs and the UE position. The APLE exhibits linear computational complexity with respect to the number of ELAA antennas and near-optimal localization performance. {However, the APLE algorithm assumes that each antenna element in the ELAA is equipped with an independent RF chain, which results in significant hardware overhead. Common methods to reduce this hardware burden include antenna selection \cite{asaad2018massive,gao2017massive,bereyhi2021detection} and analog/hybrid beamforming \cite{wu2018hybrid}. Among these, analog/hybrid beamforming is preferred in localization due to its beamforming gain. As an extension of the APLE algorithm to the analog beamforming scenario, the work in \cite{teng2024near} proposed the APLE-ABL algorithm, which further divides the subarrays into groups, simplifying the signal model by assuming a linear drift of the AoAs within each group. Nevertheless, this linear approximation introduces model errors, which degrade localization accuracy. The APLE-ABL algorithm is designed for 2D localization with a uniform linear array (ULA) and faces challenges when extended to 3D scenarios with a uniform planar array (UPA), as the AoAs have two components that drift along both the $x$-axis and $y$-axis of the subarrays. Both the APLE and APLE-ABL algorithms are limited to single-user scenarios and struggle in non-orthogonal multi-user environments due to interference from uplink signals. Furthermore, in multi-user scenarios, it is necessary for \cite{yuan2024scalable} and \cite{teng2024near} to consider the matching between the subarray AoAs and the user positions, which introduces additional challenges.}
\par
\par
In addition to \cite{yuan2024scalable,teng2024near}, signal subspace-based algorithms \cite{zhang2018localization,zhi2007near,pan2023ris,qiao2023sensing} and maximum likelihood (ML) principle-based algorithms \cite{abu2021near,chen2024elaa,lu22023near} are proposed to localize UE.
In \cite{zhang2018localization} and \cite{zhi2007near}, classical Direction-of-Arrival (DoA) estimation algorithms {multiple signal classification} (MUSIC) and Estimation of Signal Parameters via Rotational Invariance Techniques (ESPRIT) are extended to the near-field scenario. In \cite{pan2023ris}, a down-sampled Toeplitz covariance matrix is designed to decouple the estimation of range and AoA, and a subspace-based method is used to search for position estimation. A two-stage algorithm is designed in \cite{qiao2023sensing}, where the near-field channel is first estimated by using an orthogonal matching pursuit (OMP)-like algorithm, followed by UE localization based on a MUSIC-like method. In \cite{abu2021near}, the single-user 3-dimensional (3D) localization near-field localization problem is solved by obtain the ML estimator using three one-dimensional searches. The work \cite{chen2024elaa} considers a simplified 2D scenario, utilizing a wideband signal for joint localization and synchronization of a single-antenna user, while also accounting for the spatial non-stationary of the channel. The authors in \cite{lu22023near} leverage the spatial sparsity of the near-field channel to achieve joint user localization and channel estimation, designing a two-stage algorithm based on the OMP-like algorithm.
\par
{Existing approaches to near-field localization face critical limitations in practical ELAA systems.} For subspace-based near-field localization algorithms proposed in \cite{zhang2018localization,zhi2007near,pan2023ris,qiao2023sensing}, they typically require multiple observations to construct a reliable sample covariance matrix. However, in the mobile scenario, due to the fast fading of the wireless channel, it is difficult to obtain sufficient observations within the coherence time. Moreover, the array response loses its Vandermonde structure when the UE is in the near-field region, which makes the spatial smoothing method inapplicable \cite{hu2022irs}.
For the ML-based algorithms proposed in \cite{abu2021near,lu22023near,chen2024elaa}, the ELAA's high angular resolution characteristic is utilized to achieve accurate angle estimation. However, the high angular resolution also makes the corresponding ML problem highly non-convex. This issue is especially pronounced in near-field multiuser localization, where each user's position is parameterized by AoA and range, leading to a large search space and making it challenging to obtain the optimal user position estimates. 
Additionally, none of the aforementioned works \cite{zhang2018localization,
zhi2007near,
pan2023ris,
qiao2023sensing,
abu2021near,
lu22023near,
chen2024elaa,
yuan2024scalable
} accounted for the array antenna radiation pattern in the near-field channel modeling. In the far-field geometric channel model, the DoA is the same for all array antennas, leading to identical antenna gains that can be absorbed into the channel path gain. However, in the near-field channel model, the DoA varies across different array antennas, causing the directional characteristics of the antennas to distort the array response. The existing literature has yet to address how antenna radiation patterns impact near-field user localization performance.
\par
Motivated by the limitations of the existing works, we investigate the near-field localization problem for multiple single-antenna users in an uplink {MIMO} system, where the BS is equipped with an ELAA to receive signals from multiple single-antenna users and uses analog beamforming to reduce the number of RF chains. The contributions of this paper are listed as follows:
\begin{enumerate}
    \item {By incorporating the antenna radiation pattern and distance-dependent free-space path loss, we establish a comprehensive near-field ELAA channel model parameterized by user locations, thereby addressing the systematic model errors present in existing works.} We employ an array partitioning strategy to divide the near-field ELAA channel into multiple subarray channels. By leveraging the mapping between user locations and subarray channels, we establish a probabilistic model and formulate the near-field multiuser localization problem within a Bayesian framework.
    
    \item We propose an array partitioning-based location estimation with limited measurements (APLE-LM) algorithm to address the near-field multiuser localization problem based on the message-passing principle. We employ the vector sum-product rule to compute messages between variable nodes and factor nodes on the factor graph corresponding to the probabilistic model. {Gaussian approximations are employed to simplify message calculations related to the user positions. Compared to the baseline work, our approach requires significantly fewer grid search points, thereby reducing computational complexity.} 
    \item We derive the Bayesian Cram\'er–Rao Bound (BCRB) as a theoretic mean square error (MSE) lower bound for the near-field user localization and channel reconstruction. We conduct numerical simulations under various parameter configurations to validate the effectiveness of the proposed APLE-LM algorithm. We numerically show that the multiuser localization accuracy of the APLE-LM algorithm significantly outperforms the baseline schemes and requires much fewer grid search points compared to the baseline schemes. The localization performance of the proposed algorithm approaches the derived BCRB at high signal-to-noise ratio (SNR).
\end{enumerate}
\par
The remainder of this paper is organized as follows. In Section II, we introduce the near-field ELAA localization system and the signal model. In Section III, we formulate the near-field multiuser localization problem based on array partitioning approach. In Section IV, we derive the message calculations and develop the APLE-LM algorithm. In Section V, we analyze the BCRB for the considered multiuser localization problem. Numerical results are presented in Section VI. The paper concludes in Section VII.
\par
\textit{Notations:} Vectors and matrices are denoted by bold lowercase letters and bold capital letters, respectively. We use $(\cdot)^{\mathrm{T}}$, $(\cdot)^{\mathrm{H}}$, and $(\cdot)^{*}$ to denote the operations of transpose, conjugate transpose, and conjugate, respectively. We use Tr($\mathbf{X}$) to denote the trace of $\mathbf{X}$, $\Re\{\mathbf{X}\}$ and $\Im\{\mathbf{X}\}$ to denote the real part and imaginary part of $\mathbf{X}$, diag($\mathbf{x}$) to denote the diagonal matrix with its diagonal entries given by $\mathbf{x}$, {$\mathrm{blkdiag}(\mathbf{X}_1, \ldots, \mathbf{X}_n)$ to denote the block diagonal matrix formed by matrices $\mathbf{X}_1, \ldots, \mathbf{X}_n$,} and $\mathbf{I}_{n}$ to denote the $n\times n$ identity matrix, respectively. We use $\mathcal{N}\left(\boldsymbol{x};\boldsymbol{\mu},\boldsymbol{\Sigma}\right)$ and $\mathcal{CN}\left(\boldsymbol{x};\boldsymbol{\mu},\boldsymbol{\Sigma}\right)$ to denote the real Gaussian distribution and the circularly-symmetric Gaussian distribution with mean vector $\boldsymbol{\mu}$ and covariance matrix $\boldsymbol{\Sigma}$. We use $\mathbb{E}[\cdot]$ for the expectation operator, $\|\cdot\|_p$ for the $\ell_{p}$ norm, and $\jmath$ for the imaginary unit.
\begin{figure}[t]
    \centering
    \resizebox{0.9\linewidth}{!}{\includegraphics{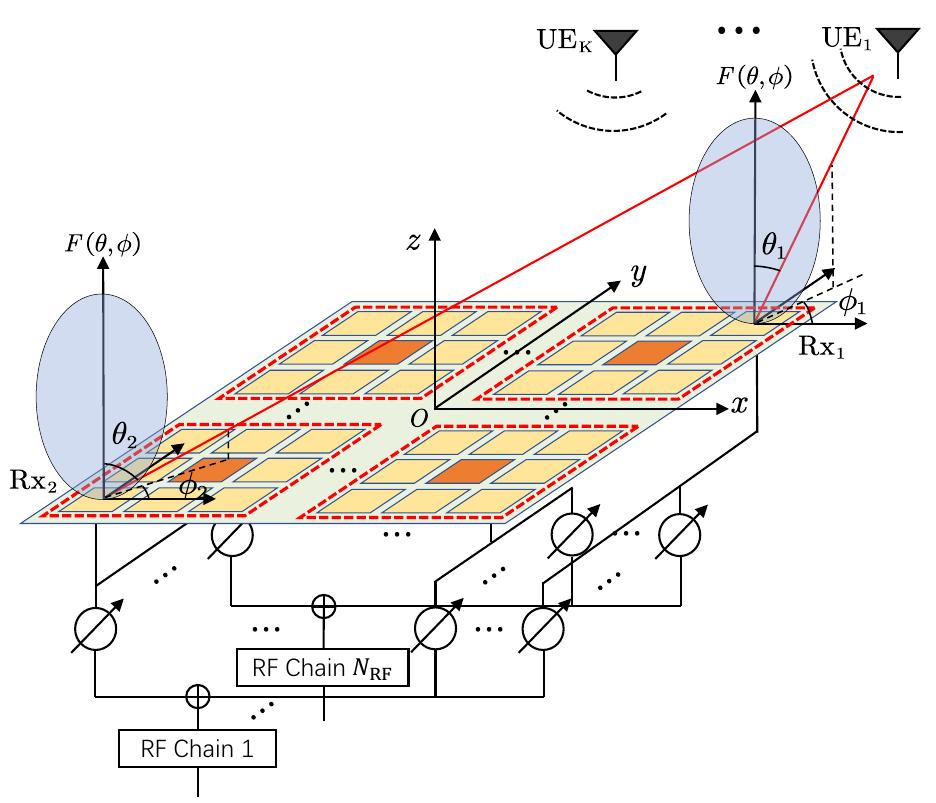}}
    \caption{System model. The BS receives the signals from $K$ single antenna users. The BS antenna array is partitioned into multiple subarrays indicated by red dotted lines. Orange antennas correspond to the reference antenna in each subarray. Rx1 and Rx2 are the top-right and bottom-left antennas, respectively. The BS is equipped with $N_{\mathrm{RF}}$ RF chains configured in a fully connected architecture.}
    \label{system}
\end{figure}
\section{System Model}\label{S2}
\subsection{System Description}
As shown in Fig. \ref{system}, we consider an uplink multiuser MIMO system, where $K$ single-antenna {UEs} transmit narrow-band signals over the same frequency and time slot to the BS. The BS is equipped with an ELAA consisting of $N_{\mathrm{B}}=N_{\mathrm{B},x}\times N_{\mathrm{B},y}$ antennas arranged in a UPA with antenna interval $d$. We assume that the BS is equipped with $N_{\mathrm{RF}}$ RF chains (with $N_{\mathrm{RF}}\ll N_{\mathrm{B}}$), to reduce the hardware cost. The fully connected analog beamforming architecture is employed where each RF chain connects to all the BS antennas using $N_{\mathrm{B}}$ phase shifters and one adder. A 3D Cartesian coordinate system is established with the origin $O$ set at the center of the BS antenna array and the $x$-axis and $y$-axes aligned with the edges of the BS array. The position of the $(i,j)$-th BS antenna is calculated by
\begin{align}
    \mathbf{q}_{i,j}=\left[\left(i-\frac{N_{\mathrm{B},x}+1}{2}\right)d,\left(j-\frac{N_{\mathrm{B},y}+1}{2}\right)d,0\right]^{\mathrm{T}}.
\end{align}
The users are located in the upper half-plane above the BS antenna array, with the position of the $k$-th user denoted by $\mathbf{p}_{k}=[\mathrm{p}_{k,x},\mathrm{p}_{k,y},\mathrm{p}_{k,z}]^{\mathrm{T}}$. The distance between the $(i,j)$-th BS antenna and the $k$-th user is calculated by 
\begin{align}\label{eqq2}
    d_{(i,j),k} = \|\mathbf{p}_{k}-\mathbf{q}_{i,j}\|_{2}.
\end{align}
\subsection{Channel Model}
{Considering the substantial aperture size of the ELAA, UEs are assumed to be localized in the near-field region of the BS array \cite{cui2022near}. We develop a comprehensive channel model between the UE and the ELAA, incorporating both the near-field spherical wavefront and the precise distances between corresponding transmitting and receiving antenna elements. The proposed channel model accounts for two critical factors: the directional antenna radiation pattern and the free-space path loss effect.}
The antenna power radiation pattern describes the angular variation of radiation level around an antenna. We assume that the users are equipped with an isotropic antenna that has uniform radiation in all directions. Each antenna in the ELAA has the normalized power radiation pattern given by \cite{tang2020wireless,stutzman2012antenna}
\begin{align}\label{eq3}
    F(\theta ,\varphi )=\begin{cases}
	\cos ^3\theta&		\theta \in \left[ 0,\frac{\pi}{2} \right] ,\varphi \in [0,2\pi ]\\
	0&		\theta \in \left( \frac{\pi}{2},\pi \right] ,\varphi \in [0,2\pi ]
\end{cases},
\end{align}
where $\theta$ and $\varphi$ are the elevation and azimuth angles of a specific incoming plane wave direction.\footnote{The radiation pattern of an antenna is determined by various factors, including antenna geometry, antenna polarization, and antenna material. The algorithm developed in this paper can be straightforwardly applied to any other radiation pattern.} In Fig. \ref{system}, Rx1 and Rx2 are two receiving antennas far apart on the BS antenna array. The normalized radiated power of the wave direction from UE1 to Rx1 is stronger than that to Rx2 since the plane wave to Rx1 has a smaller elevation angle $\theta_{1} < \theta_{2}$. Such a radiated power difference exists across the entire array.
\par
The free space path loss is determined by the Friis transmission equation. The ratio of received signal power $P_r$ at the $(i,j)$-th BS antenna to the power transmitted $P_t$ at the $k$-th user antenna is\cite{volakis2007antenna}
\begin{align}\label{eq4}
    \frac{P_r}{P_t}=\left(\frac{\lambda}{4\pi d_{(i,j),k}}\right)^2.
\end{align}
Based on \eqref{eq3} and \eqref{eq4}, the channel coefficient between the $k$-th user and the $(i,j)$-th BS antenna is given by
\begin{align}
\label{eq5}
    h_{(i,j),k} = \alpha_{k}\gamma_{(i,j),k}\frac{1}{d_{(i,j),k}}\exp\left({-\jmath2\pi\frac{d_{(i,j),k}}{\lambda}}\right),
\end{align}
where $\frac{1}{d_{(i,j),k}}$ accounts for the free space path loss, $\gamma_{(i,j),k}$ accounts for the antenna power radiation in \eqref{eq3} and is parameterized by user position $\mathbf{p}_{k}$ as
\begin{align}
    \gamma_{(i,j),k}=\left(\frac{\mathbf{e}_{z}^{\mathrm{T}}(\mathbf{p}_{k}-\mathbf{q}_{i,j})}{\|\mathbf{p}_{k}-\mathbf{q}_{i,j}\|_{2}}\right)^{\frac{3}{2}}
\end{align}
with $\mathbf{e}_{z}=[0,0,1]^{\mathrm{T}}$, and
{$\alpha_{k}$ represents the complex channel gain, which incorporates factors unrelated to the UE positions, such as a constant terms, transmitter antenna gain, rain attenuation, and other influences. For simplicity, this paper considers only the constant terms and set $\alpha_{k} = \frac{\lambda^2}{16\pi^2}$.}
By stacking all the channel elements, we obtain the channel between the $k$-th user and the $N_{\mathrm{B},x}\times N_{\mathrm{B},y}$ BS antennas as
\begin{align}\label{eq8}
    \boldsymbol{H}_k=\left[ \begin{matrix}
	h_{\left( 1,1 \right) ,k}&		\cdots&		h_{\left( 1,N_{\mathrm{B},y} \right) ,k}\\
	\vdots&		\ddots&		\vdots\\
	h_{\left( N_{\mathrm{B},x},1 \right) ,k}&		\cdots&		h_{\left( N_{\mathrm{B},x},N_{\mathrm{B},y} \right) ,k}\\
\end{matrix} \right] \in \mathbb{C} ^{N_{\mathrm{B},x}\times N_{\mathrm{B},y}}.
\end{align}
The corresponding channel vector is given by $\boldsymbol{h}_{k}=vec(\boldsymbol{H}_k)$, which is parameterized by the user positions $\mathbf{p}_{k}$ and the complex channel gains $\alpha_{k}$.
\begin{remark}
    We assign the $\left(\left\lceil\frac{N_{\mathrm{B},x}+1}{2}\right\rceil,\left\lceil\frac{N_{\mathrm{B},y}+1}{2}\right\rceil\right)$-th BS antenna as the reference antenna with the channel coefficient denoted by $h_{\mathrm{ref},k}=\alpha_{k}\gamma_{\mathrm{ref},k}\frac{1}{d_{\mathrm{ref},k}}\exp\left({-\jmath2\pi\frac{d_{\mathrm{ref},k}}{\lambda}}\right)$. The channel model in \cite{zhang2023near,lu2023near,cui2022channel,pan2023ris} is an approximation of \eqref{eq8} by assuming 
\begin{align}\label{eq11}
    \frac{\gamma_{(i,j),k}}{\gamma_{\mathrm{ref},k}}=1\ \mathrm{and}\ \frac{d_{\mathrm{ref},k}}{d_{(i,j),k}}=1
\end{align}
for $1\leq i\leq N_{\mathrm{B},x}$ and $1\leq j\leq N_{\mathrm{B},y}$ {with the $(i,j)$-th term of $\boldsymbol{H}_{k}$ given by $[\boldsymbol{H}_{k}]_{i,j}=h_{\mathrm{ref},k}\exp\left({-\jmath2\pi\frac{(d_{(i,j),k}-d_{\mathrm{ref},k})}{\lambda}}\right)$.}
{
\begin{remark}
By applying the far-field plane wavefront approximation, we have $d_{(i,j),k}-d_{(i',j'),k}=(i-i')d\cos(\vartheta_{k,x})+(j-j')d\cos(\vartheta_{k,y})$, where $\vartheta_{k,x}^{(u,v)}$ and $\vartheta_{k,y}^{(u,v)}$ are the AoAs of the $k$-th user along the $x$-axis and $y$-axis of the BS antenna array, respectively. Together with the approximation in \eqref{eq11}, the channel in \eqref{eq8} degenerates to the far-field channel in \cite{teng2022bayesian,wang2021joint}, expressed by
\begin{align}
    \label{eq17}\boldsymbol{H}_{k}=h_{(1,1),k}\mathbf{a}_{N_{\mathrm{B},x}}\left(\vartheta_{k,x}\right)\mathbf{a}_{N_{\mathrm{B},y}}^{\mathrm{T}}\left(\vartheta_{k,y}\right),
\end{align}
where $\mathbf{a}_{N}(\vartheta)=[1,e^{\jmath2\pi\frac{d}{\lambda}\cos(\vartheta)}...,e^{\jmath2\pi\frac{d}{\lambda}(N-1)\cos(\vartheta)}]^{\mathrm{T}}$ is the steering vector.
\end{remark}}
\end{remark}
\subsection{Signal Model}
Given the channel model \eqref{eq8}, the received signal is expressed as
\begin{align}\label{eq12nn}
    \boldsymbol{y}=\mathbf{W}\sum_{k=1}^{K}\sqrt{P_{k}}\boldsymbol{h}_{k}x_{k}+\boldsymbol{n},
\end{align}
where $x_{k}$ denotes the transmitted signal from the $k$-th user, $P_{k}$ is the transmitted power of the $k$-th user, $\mathbf{W}\in\mathbb{C}^{N_{\mathrm{RF}}\times N_{\mathrm{B}}}$ represents the analog beamforming matrix satisfying the unit-modulus constraint $|\left[\mathbf{W}\right]_{i,j}|=1$, and $\boldsymbol{n}$ is the complex Gaussian noise following $\mathcal{CN}(\boldsymbol{n};\mathbf{0},\boldsymbol{C}_{n})$ with $\boldsymbol{C}_{n}=\sigma_{n}^{2}\mathbf{W}\mathbf{W}^{\mathrm{H}}$. The transmit signal $x_{k}$ can be communication data or pilot signals.
Letting $\beta_{k}=\sqrt{P_{k}}x_{k}\alpha_{k}$, we express \eqref{eq12nn} as
\begin{align}
    \boldsymbol{y}=\mathbf{W}\sum_{k=1}^{K}\beta_{k}\mathbf{a}(\mathbf{p}_{k})+\boldsymbol{n},
\end{align}
where $\mathbf{a}(\mathbf{p}_{k})$ is referred to as the near-field array response vector with the $i$-th entry given by $[\mathbf{a}(\mathbf{p}_{k})]_{i} = \frac{[\boldsymbol{h}_{k}]_{i}}{\alpha_{k}}$.
\par
The BS estimates the user positions from the received signal $\boldsymbol{y}$. A straightforward method is to exploit the ML principle for user position estimation.
Denote by $\hat{\boldsymbol{\beta}}=(\mathbf{A}^{\mathrm{H}}\boldsymbol{C}_{n}^{-1}\mathbf{A})^{-1}\mathbf{A}^{\mathrm{H}}\boldsymbol{C}_{n}^{-1}\boldsymbol{y}$ the least-square estimate of $\boldsymbol{\beta}=[\beta_{1},...,\beta_{K}]^{\mathrm{T}}$, where $\mathbf{A}=\left[\mathbf{W}\mathbf{a}(\mathbf{p}_{1}),...,\mathbf{W}\mathbf{a}(\mathbf{p}_{K})\right]$.
The ML estimation is obtained by solving
\begin{align}\label{eq13}
    \underset{\{\mathbf{p}_{k}\}_{k=1}^{K}}{\max}\ \ln p\left(\boldsymbol{y};\{\mathbf{p}_{k}\}_{k=1}^{K},\hat{\boldsymbol{\beta}}\right).
\end{align}
As illustrated later in Fig. \ref{fig:4}(a), the log-likelihood function $\ln p\left(\boldsymbol{y};\{\mathbf{p}_{k}\}_{k=1}^{K},\hat{\boldsymbol{\beta}}\right)$ is highly non-convex with a large number of local maxima.
Gradient ascent methods can easily get stuck in these local maxima, posing challenges in finding the global optimal solution. Consequently, a fine-grid search becomes essential to escape from local maxima. However, performing such an exhaustive search in a 3D space can be extremely time-consuming.
\par
To avoid the computational complexity of the fine-grid search, the prior work \cite{yuan2024scalable} proposes the array partitioning-based location estimation (APLE) algorithm, where the ELAA is partitioned into multiple subarrays. The AoAs at each subarray are estimated using the far-field channel model and are then used to estimate the user’s position based on geometric constraints. However, when the number of radio frequency chains of the BS is limited, the signals of each subarray are superimposed through analog beamforming, making the APLE algorithm in \cite{yuan2024scalable} not directly applicable. It is difficult to estimate the AoA at each subarray from the signal mixture since the AoAs of adjacent subarrays differ only slightly. In this paper, we estimate the user’s position directly without the need to explicitly estimate the AoAs of the subarrays. By introducing array partitioning to model the received signal, we can avoid directly handling the highly non-convex user position estimation problem, as explained in detail later. Within the Bayesian framework, the geometric relationships between different subarrays are further leveraged to enhance position estimation performance.
\section{Array Partitioning and Probabilistic Problem Formulation}
\subsection{Array Partitioning} The BS antenna array is evenly partitioned into $M\times M$ subarrays with each subarray being a rectangle consisting of ${N_{\mathrm{S},x}}\times{N_{\mathrm{S},y}}$ antennas with $N_{\mathrm{S},x}=\frac{N_{\mathrm{B},x}}{M}$ and $N_{\mathrm{S},y}=\frac{N_{\mathrm{B},y}}{M}$. 
The matrix form of the channel between the $k$-th user and the $(u,v)$-th subarray is denoted by $\boldsymbol{H}_{k}^{(u,v)}$, which is a submatrix of $\boldsymbol{H}_{k}$ with the row indexed by $(u-1)N_{\mathrm{S},x}+1:uN_{\mathrm{S},x}$ and column indexed by $(v-1)N_{\mathrm{S},y}+1:vN_{\mathrm{S},y}$. 
The $\left((u-1)N_{\mathrm{S},x}+\left\lceil\frac{N_{\mathrm{S},x}}{2}\right\rceil,(v-1)N_{\mathrm{S},y}+\left\lceil\frac{N_{\mathrm{S},y}}{2}\right\rceil\right)$-th BS antenna is assigned as the reference antenna of the $(u,v)$-th subarray. Denote by $h_{ref,k}^{(u,v)}$ the channel coefficient between the reference antenna of the $(u,v)$-th subarray and the $k$-th user. We express $\boldsymbol{H}_{k}^{(u,v)}$ as
\begin{align}\label{eq15}
    \boldsymbol{H}_{k}^{(u,v)}=h_{ref,k}^{(u,v)}\mathbf{D}^{(u,v)}(\mathbf{p}_{k}),
\end{align}
where the $(i,j)$-th term of $\mathbf{D}^{(u,v)}(\mathbf{p}_{k})$ is calculated by $[\mathbf{D}^{(u,v)}(\mathbf{p}_{k})]_{i,j}=\frac{[\boldsymbol{H}_{k}^{(u,v)}]_{i,j}}{h_{ref,k}^{(u,v)}}$. {In this work, we adopt the rectangular array partitioning approach for notational simplicity. It is worth noting that the array partitioning strategy is not unique. For different partitioning strategies, only minor modifications to the definition of $\mathbf{D}^{(u,v)}(\mathbf{p}_{k})$ in \eqref{eq15} are required to match the specific partitioning form. The APLE-LM algorithm proposed in Section IV is flexible and can be applied to various array partitioning strategies.}
\par
%
\subsection{Array-Partitioning-Based Signal Model}
There are geometric constraints between the reference antennas of different subarrays, i.e., the reference coefficients of different subarrays $h_{ref,k}^{(u,v)}$ for $1\leq u,v \leq M$  can be parameterized by the user position {$\mathbf{p}_{k}$}. 
{Specifically, we collect all the channel coefficients between the reference antennas and the $k$-th UE as $\boldsymbol{h}_{ref,k}\in\mathbb{C}^{M^{2}}$, which is given by
\begin{subequations}
    \begin{align}
    \boldsymbol{h}_{ref,k}&=\left[h_{ref,k}^{(1,1)},...,h_{ref,k}^{(M,1)},...,h_{ref,k}^{(1,M)},...,h_{ref,k}^{(M,M)}\right]^\mathrm{T}\\
    &=h_{ref,k}^{(\frac{M}{2},\frac{M}{2})}\mathbf{c}(\mathbf{p}_{k}),
\end{align}
\end{subequations}}
where the $s$-th item of $\mathbf{c}(\mathbf{p}_{k})$ is given by
\begin{align}\label{eq18}
    \left[\mathbf{c}(\mathbf{p}_{k})\right]_{s}=\frac{\gamma_{ref,k}^{(u,v)}}{\gamma_{ref,k}^{(\frac{M}{2},\frac{M}{2})}}{\frac{d_{ref,k}^{(\frac{M}{2},\frac{M}{2})}}{d_{ref,k}^{(u,v)}}e^{-\frac{\jmath2\pi}{\lambda}\left(d_{ref,k}^{(u,v)}-d_{ref,k}^{(\frac{M}{2},\frac{M}{2})}\right)}},
\end{align}
with the index mapping from $s$ to $(u,v)$ given by $s=(v-1)M+u$. We reconstruct $\boldsymbol{H}_{k}$ by using $h_{ref,k}^{(\frac{M}{2},\frac{M}{2})}$, $\mathbf{c}(\mathbf{p}_{k})$, and $\mathbf{D}^{(u,v)}(\mathbf{p}_{k})$ for $1\leq u,v\leq M$ as
\begin{align}\label{eq20}
    \boldsymbol{H}_k=h_{ref,k}^{(\frac{M}{2},\frac{M}{2})}\left[ \begin{matrix}
	\left[ \mathbf{c}_{k} \right] _1\mathbf{D}_{k}^{(1,1)}&		\cdots&		\left[ \mathbf{c}_{k} \right] _{\left( M-1 \right) M+1}\mathbf{D}_{k}^{(1,M
 )}\\
	\vdots&		\ddots&		\vdots\\
	\left[ \mathbf{c}_{k} \right] _M\mathbf{D}_{k}^{(M,1)}&		\cdots&		\left[ \mathbf{c}_{k} \right] _{M^2}\mathbf{D}_{k}^{(M,M)}\\
\end{matrix} \right],
\end{align}
where $\mathbf{c}_{k}$ and $\mathbf{D}_{k}^{(u,v)}$ respectively denote $\mathbf{c}(\mathbf{p}_{k})$ and $\mathbf{D}^{(u,v)}(\mathbf{p}_{k})$ for brevity. 
Based on \eqref{eq20}, the received signal in \eqref{eq12nn} is further expressed by
\begin{align}
    \label{eq9}\boldsymbol{y}=\sum_{k=1}^{K}\varrho_{k}\sum_{s=1}^{M^{2}}[\mathbf{c}_{k}]_{s}\mathbf{W}_{s}\mathbf{d}_{k,s}+\boldsymbol{n},
\end{align}
where $\varrho_{k}=\sqrt{P_{k}}x_{k}h_{ref,k}^{(\frac{M}{2},\frac{M}{2})}$ is the reference array gain of user $k$, $\mathbf{W}_{s}\in\mathbb{C}^{N_{\mathrm{RF}}\times N_{\mathrm{S},x}N_{\mathrm{S},y}}$ is the analog beamforming matrix corresponding to the $s$-th subarray, and $\mathbf{d}_{k,s}=vec(\mathbf{D}_{k}^{(u,v)})$. We define auxiliary variables
\begin{subequations}
    \begin{align}
    \boldsymbol{\rho}_{k}&=\varrho_{k}\mathbf{c}_{k}\in\mathbb{C}^{M^2},\ \boldsymbol{\rho}=\left[\boldsymbol{\rho}_{1}^{\mathrm{T}},...,\boldsymbol{\rho}_{K}^{\mathrm{T}}\right]^{\mathrm{T}}\in\mathbb{C}^{KM^2},\label{eq21a}\\
    \mathbf{B}_{k}&=\left[\mathbf{W}_{1}\mathbf{d}_{k,1},...,\mathbf{W}_{M^2}\mathbf{d}_{k,M^2}\right]\in\mathbb{C}^{N_{\mathrm{RF}}\times M^2},\label{eq21b}\\
    \mathbf{B}& = \left[\mathbf{B}_{1},...,\mathbf{B}_{K}\right]\label{eq21c}\in\mathbb{C}^{N_{\mathrm{RF}}\times KM^2}.
\end{align}
\end{subequations}
A compact form of \eqref{eq9} is given by 
\begin{align}
    \label{eq23}
    \boldsymbol{y}=\mathbf{B}\boldsymbol{\rho}+\boldsymbol{n}.
\end{align}
{In the array-partitioning-based signal model \eqref{eq23}, we introduce the variable $\boldsymbol{\rho}$, which allows us to obtain the position information from both the geometric constraints between $\mathbf{B}$ and the user position, and the geometric constraints between $\boldsymbol{\rho}$ and the user position, rather than directly estimating position from $\boldsymbol{y}$ as in \eqref{eq13}.}
In this paper, we develop a positioning algorithm based on the array-partitioning-based signal model \eqref{eq23}.
\subsection{Probabilistic Problem Formulation}\label{III-C}
We assign a complex Gaussian prior distribution to $\varrho_{k}$ and a Gaussian prior distribution to $\mathbf{p}_{k}$ as given by
\begin{subequations}
\label{eq24}
    \begin{align}
    p(\varrho_{k})&=\mathcal{CN}\left(\varrho_{k};0,\tau_{k}^{\mathrm{pri}}\right),\label{eq24a}\\
    p(\mathbf{p}_{k})&=\mathcal{N}\left(\mathbf{p}_{k};\boldsymbol{\mu}_{\mathbf{p}_k}^{\mathrm{pri}},\nu_{k}^{\mathrm{pri}}\mathbf{I}_{3}\right).\label{eq24b}
\end{align}
\end{subequations}
If the prior knowledge on $\mathbf{p}_{k}$ and $\varrho_{k}$ is unavailable, $\tau_{k}^{\mathrm{pri}}$ and $\nu_{k}^{\mathrm{pri}}$ are set to a relatively large positive number, making \eqref{eq24} non-informative prior distributions. Given the equation in \eqref{eq21a} and noting that $\mathbf{c}_{k}$ is parameterized by $\mathbf{p}_{k}$ as shown in \eqref{eq18}, we have
\begin{subequations}\label{7_17_eq19}
    \begin{align}
    p(\boldsymbol{\rho}|\mathbf{p},\boldsymbol{\varrho})&=\prod_{k=1}^{K}p(\boldsymbol{\rho}_{k}|\mathbf{p}_{k},\varrho_{k})\\
    &=\prod_{k=1}^{K}\delta\left(\boldsymbol{\rho}_{k}-\varrho_{k}\mathbf{c}_{k}\right),
\end{align}
\end{subequations}
where $\mathbf{p}=[\mathbf{p}_{1}^{\mathrm{T}},...,\mathbf{p}_{K}^{\mathrm{T}}]^{\mathrm{T}}$,
$\boldsymbol{\varrho}=[\varrho_{1},...,\varrho_{K}]^{\mathrm{T}}$ and
$\delta(\cdot)$ is the Dirac delta function.
{$p(\boldsymbol{\rho}|\mathbf{p},\boldsymbol{\varrho})$ in \eqref{7_17_eq19} represents the geometric constraints between the reference antennas of different subarrays and the user position.}
Based on the observation model in \eqref{eq23} together with the fact that $\mathbf{B}$ is parameterized by $\mathbf{p}$, we obtain
\begin{align}
    \label{eqq25}p(\boldsymbol{y}|\mathbf{p},\boldsymbol{\rho})=\mathcal{CN}\left(\boldsymbol{y};\mathbf{B}\boldsymbol{\rho},\boldsymbol{C}_{n}\right)
\end{align}
{which represents the geometric constraint between the signals within a subarray and the user position.}
The joint probability density function (pdf) is given by
\begin{align}\label{eq27}
    p\left( \boldsymbol{y},\boldsymbol{\rho },\mathbf{p},\boldsymbol{\varrho} \right) =
    p\left( \boldsymbol{y}|\mathbf{p}, \boldsymbol{\rho }\right) p\left( \boldsymbol{\rho }|\mathbf{p},\boldsymbol{\varrho}\right)p\left( \boldsymbol{\varrho} \right)p\left( \mathbf{p} \right),
\end{align}
where $p\left( \boldsymbol{\varrho} \right)=\prod_{k=1}^{K}p\left( \varrho _k \right)$ and $p\left(\mathbf{p} \right)=\prod_{k=1}^K{  p\left( \mathbf{p}_k \right)}$. Following the Bayes' theorem, the posterior pdf of $\mathbf{p}_k$ is calculate by
\begin{align}\label{eq28}
    p(\mathbf{p}_k|\boldsymbol{y})=\int_{\backslash\mathbf{p}_k}\frac{p\left( \boldsymbol{y},\boldsymbol{\rho },\mathbf{p},\boldsymbol{\varrho} \right)}{p\left( \boldsymbol{y}\right)},
\end{align}
where $\int_{\backslash \mathbf{p}_{k}}$ represents integral over $\{\mathbf{p}_k,\boldsymbol{\rho }_k,\varrho _k\}_{k=1}^{K}$ except $\mathbf{p}_{k}$. The traditional minimum mean-square error (MMSE) and maximum \textit{a posteriori} (MAP) estimators are generally intractable due to the high-dimensional integrals in \eqref{eq28}. In the following, we provide a near-optimal approximate solution by following the message-passing principle.
\begin{figure}[t]
    \centering
    \resizebox{0.65\linewidth}{!}{\includegraphics{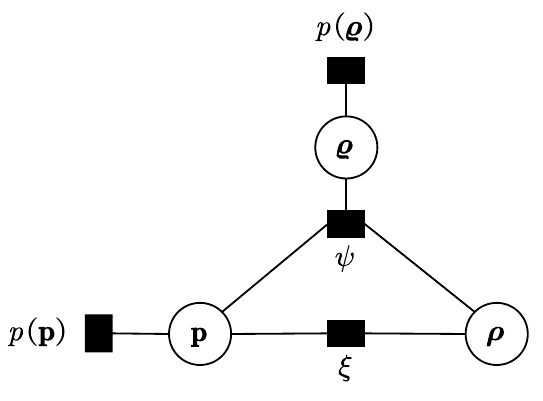}}
    \caption{The factor graph representation corresponding to the joint pdf in \eqref{eq27}. 
    }
    \label{factor}
\end{figure}
\begin{figure*}[htb]
\centering
\subfloat[]{
        \includegraphics[width=0.32\linewidth,height=4.6 cm]{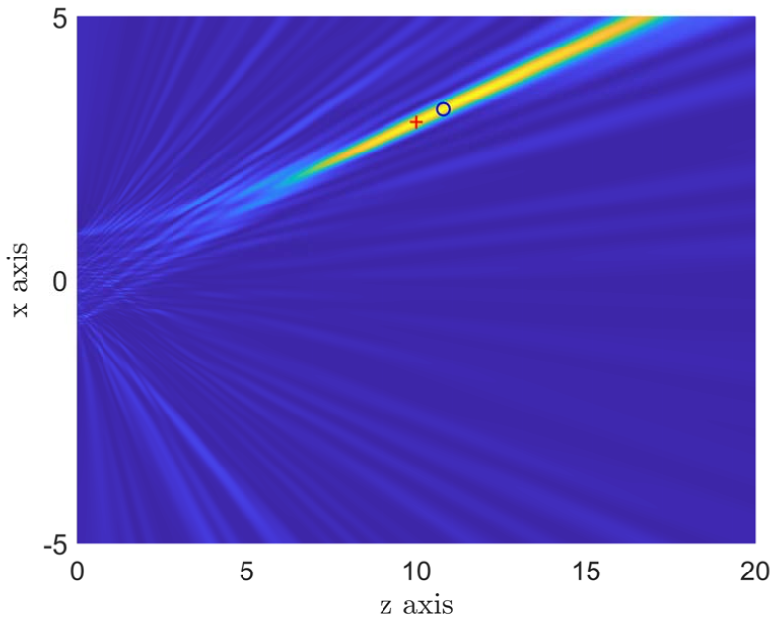}
        \label{fig:4a}
    }
    \subfloat[]{\centering
        \includegraphics[width=0.32\linewidth,height=4.6 cm]{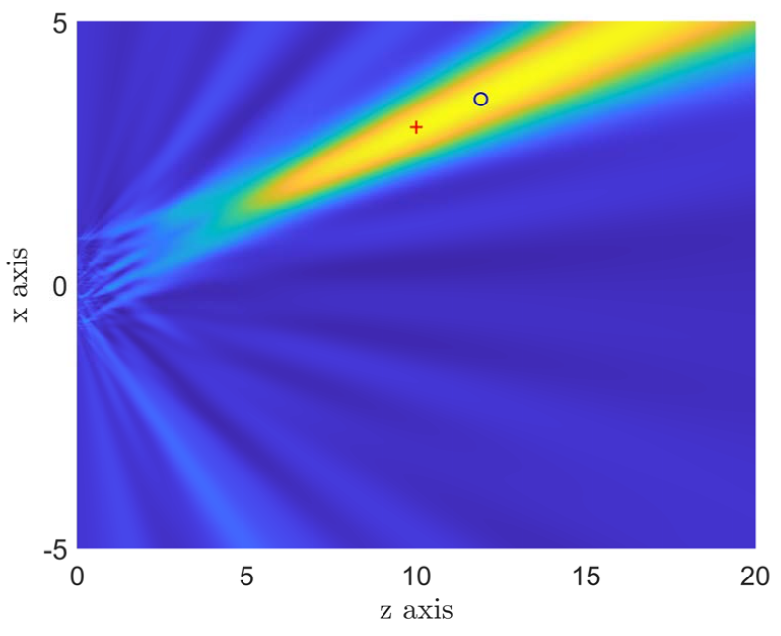}
        \label{fig:4b}
    }
    \subfloat[]{\centering
        \includegraphics[width=0.32\linewidth,height=4.6 cm]{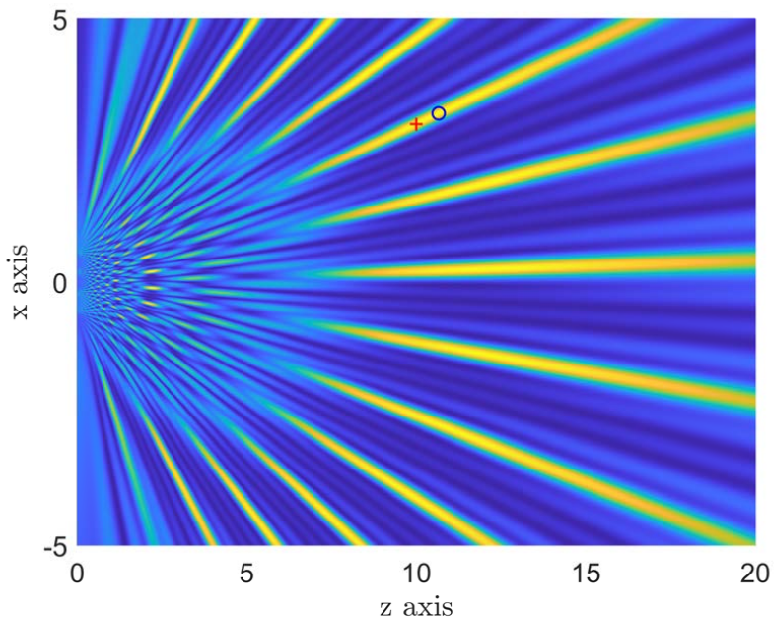}
        \label{fig:4c}
    }
    \caption{Illustrations of distributions in \eqref{eq13}, \eqref{eq31}, and \eqref{eqq42}. We consider a single-user 2D positioning scenario where the user is located in the $z-x$ plane. The BS is equipped with $N_{\mathrm{B}}=75\times1$ antennas. The BS antenna array is partitioned into $5$ subarrays along the $x$-axis. The marker ``+'' denotes the user's true position and ``o'' represents the maxima of the corresponding distributions.  (a) The distribution of log-likelihood function in \eqref{eq13}; (b) The distribution of $\ln \Delta_{\xi\rightarrow\mathbf{p}}(\mathbf{p})$ with $\Delta_{\xi\rightarrow\mathbf{p}}(\mathbf{p})$ given in \eqref{eq31}; (c) The distribution of $\ln \Delta_{\psi\rightarrow\mathbf{p}}(\mathbf{p})$ with $\Delta_{\psi\rightarrow\mathbf{p}}(\mathbf{p})$ given in \eqref{eqq42}.}
    \label{fig:4}
\end{figure*}
\section{Proposed APLE-LM Algorithm}\label{S5}
In this section, we introduce the proposed message-passing-based APLE-LM algorithm. We employ the \textit{vector} sum-product rule to calculate messages between vector-form variable nodes \cite{zhang2021unifying,jiang2024hybrid}. The advantage of the \textit{vector} sum-product rule is that the messages can capture the correlations among the elements within a vector variable, leading to improved convergence performance in the message-passing algorithm.
The factor graph corresponding to the joint pdf
of \eqref{eq27} is shown in Fig. \ref{factor}, where the variable nodes and the factor nodes are denoted by blank circles and black rectangles, respectively. For notation brevity, we use $\xi$ and $\psi$ to represent factors $p\left( \boldsymbol{y}|\mathbf{p},\boldsymbol{\rho } \right)$ and $p(\boldsymbol{\rho}|\mathbf{p},\boldsymbol{\varrho})$, respectively. Denote by $\Delta_{a\rightarrow b}$ the message passing from node $a$ to $b$ with the mean $\boldsymbol{\mu}_{a\rightarrow b}$ and covariance $\boldsymbol{C}_{a\rightarrow b}$, and by $\Delta_{a}$ the marginal message at variable node $a$ with the mean $\boldsymbol{\mu}_{a}$ and covariance $\boldsymbol{C}_{a}$. We iteratively calculate the messages in Fig. \ref{factor} to estimate the user positions $\{\mathbf{p}_{k}\}_{k=1}^{K}$ by following the sum-product rule. {To simplify message calculation, the Gaussian approximation is used to compute the messages related to the user positions.}
\subsection{Calculation of Messages at Node $\xi$}
\subsubsection{Messages between $\xi$ and $\mathbf{p}$}The message passing from $\xi$ to $\mathbf{p}$ is calculated by
\begin{align}\label{eq29}
    \Delta_{\xi\rightarrow\mathbf{p}}(\mathbf{p})\propto\int_{\boldsymbol{\rho}}p\left( \boldsymbol{y}|\mathbf{p},\boldsymbol{\rho } \right)\Delta_{\boldsymbol{\rho}\rightarrow\xi}(\boldsymbol{\rho}),
\end{align}
where $p\left( \boldsymbol{y}|\mathbf{p},\boldsymbol{\rho } \right)$ {is the geometric constrain in \eqref{eqq25}} and $\Delta_{\boldsymbol{\rho}\rightarrow\xi}(\boldsymbol{\rho})$ is given by \eqref{eq34}. 
Calculate the integral in \eqref{eq29} we obtain
\begin{align}\label{eq31}
    \Delta_{\xi\rightarrow\mathbf{p}}(\mathbf{p})\propto\exp\left(\boldsymbol{\varXi}_{\xi}^{\mathrm{H}}\boldsymbol{\varUpsilon}_{\xi}^{-1}\boldsymbol{\varXi}_{\xi}\right),
\end{align}
where $\boldsymbol{\varXi}_{\xi} = \mathbf{B}^{\mathrm{H}}\boldsymbol{C}_{n}^{-1}\boldsymbol{y}$ and  $\boldsymbol{\varUpsilon}_{\xi} = \boldsymbol{C}_{\boldsymbol{\rho}\rightarrow\xi}^{-1}+\mathbf{B}^{\mathrm{H}}\boldsymbol{C}_{n}^{-1}\mathbf{B}$. {Although equation (25) provides the closed-form expression for \( \Delta_{\xi \rightarrow \mathbf{p}}(\mathbf{p}) \), the mapping of matrix \( \mathbf{B}_{k} \) with respect to the user position \( \mathbf{p}_{k} \) is highly nonlinear. Consequently, \( \Delta_{\xi \rightarrow \mathbf{p}}(\mathbf{p}) \) becomes a complex probability density function with respect to the position, making it challenging to compute subsequent messages.} To facilitate subsequent message calculations, the message in \eqref{eq31} is approximated by a Gaussian distribution
\begin{align}\label{eq32}
    \Delta_{\xi\rightarrow\mathbf{p}}(\mathbf{p})\propto\mathcal{N}(\mathbf{p};\boldsymbol{\mu}_{\xi\rightarrow\mathbf{p}},\boldsymbol{C}_{\xi\rightarrow\mathbf{p}}).
\end{align}
The mean $\boldsymbol{\mu}_{\xi\rightarrow\mathbf{p}}$ is obtained by searching the maximum of $\ln\Delta_{\xi\rightarrow\mathbf{p}}(\mathbf{p})$ as
\begin{align}\label{eq33}
    \boldsymbol{\mu}_{\xi\rightarrow\mathbf{p}}=\mathrm{arg}\ \underset{\mathbf{p}}{\max}\ \ln\Delta_{\xi\rightarrow\mathbf{p}}(\mathbf{p}),
\end{align}
{where $\Delta_{\xi\rightarrow\mathbf{p}}(\mathbf{p})$ is given by \eqref{eq31}.}
As illustrated in Fig. \ref{fig:4}(b), compared to the objective function in \eqref{eq13} shown in Fig. \ref{fig:4}(a), the global optimum peak of the objective function in \eqref{eq33} {exhibits a broader main peak. The broader peak reduces sensitivity to initialization, thereby decreasing the computational effort required for grid search to locate a starting point near the global optimum and mitigating the risk of convergence to local maxima associated with secondary peaks.}
Since the objective function in \eqref{eq33} is non-convex and $\boldsymbol{\mu}_{\xi\rightarrow\mathbf{p}}$ generally lacks a closed-form solution, we employ the \textit{gradient ascent} (GA) method and \textit{alternating optimization} (AO) to find the local maximum.
Specifically, we represent the $k$-th user position in the polar domain as given by
\begin{align}\label{eqq32n}
    \mathbf{p}_{k}=\left[r_{k}\chi_{k,x},r_{k}\chi_{k,y},r_{k}\sqrt{1-\chi_{k,x}^{2}-\chi_{k,y}^{2}}\right]^{\mathrm{T}},
\end{align}
where $r_{k}$ is the distance between the $k$-th user and the center of the BS antenna array, $\chi_{k,x}$ (or $\chi_{k,y}$) is the cosine value of the angle between the user direction and the $x$-axis (or $y$-axis). 
The problem in \eqref{eq33} is equivalent to
\begin{align}\label{eq33n}
    (\boldsymbol{\chi}^{*},\boldsymbol{r}^{*})=\mathrm{arg}\ \underset{\boldsymbol{\chi},\boldsymbol{r}}{\max}\ \ln\Delta_{\xi\rightarrow\mathbf{p}}(\boldsymbol{\chi},\boldsymbol{r}),
\end{align}
where $\boldsymbol{\chi}=\left[\chi_{1,x},\chi_{1,y},...,\chi_{K,x},\chi_{K,y}\right]^{\mathrm{T}}$ and $\boldsymbol{r}=\left[r_{1},...,r_{K}\right]^{\mathrm{T}}$.
In the $(n+1)$-th gradient ascent iteration, 
$\boldsymbol{\chi}$ and $\boldsymbol{r}$
are alternatively updated by
\begin{subequations}\label{eqq32}
    \begin{align}
    \boldsymbol{\chi}^{(n+1)}=\boldsymbol{\chi}^{(n)}+\iota_{\boldsymbol{\chi}}^{(n)}\left.\frac{\partial\ln\Delta_{\xi\rightarrow\mathbf{p}}(\boldsymbol{\chi},\boldsymbol{r})}{\partial\boldsymbol{\chi}}\right|_{(\boldsymbol{\chi},\boldsymbol{r})= (\boldsymbol{\chi}^{(n)},\boldsymbol{r}^{(n)})},\\
    \boldsymbol{r}^{(n+1)}=\boldsymbol{r}^{(n)}+\iota_{\boldsymbol{r}}^{(n)}\left.\frac{\partial\ln\Delta_{\xi\rightarrow\mathbf{p}}(\boldsymbol{\chi},\boldsymbol{r})}{\partial\boldsymbol{r}}\right|_{(\boldsymbol{\chi},\boldsymbol{r})= (\boldsymbol{\chi}^{(n)},\boldsymbol{r}^{(n)})},
\end{align}
\end{subequations}
where $\iota_{\boldsymbol{\chi}}^{(n)}$ and $\iota_{\boldsymbol{r}}^{(n)}$ are the step sizes determined by Armijo backtracking line search.\footnote{{The Armijo backtracking line search is an iterative optimization method for determining a step size that ensures sufficient descent in the objective function. It adaptively reduces the step size until the Armijo condition is met, ensuring both a sufficient reduction in the objective function and an appropriately chosen step size \cite{boyd2004convex}.}}
The gradient ascent starts from $(\boldsymbol{\chi}^{*}, \boldsymbol{r}^{*})$ given by the message calculation in the previous message iteration and stops when $\|\boldsymbol{r}^{(n+1)} - \boldsymbol{r}^{(n)}\| < \epsilon$, where $\epsilon$ is a threshold parameter.
Then,
$\boldsymbol{\mu}_{\xi\rightarrow\mathbf{p}}$ is obtained by
\begin{align}\label{eqq35n}
    \boldsymbol{\mu}_{\xi\rightarrow\mathbf{p}}=\left[f^{\mathrm{T}}(\chi_{1,x}^{*},\chi_{1,y}^{*},{r}_{1}^{*}),...,f^{\mathrm{T}}(\chi_{K,x}^{*},\chi_{K,y}^{*},{r}_{K}^{*})\right]^{\mathrm{T}},
\end{align}
where $\mathbf{p}_{k}=f(\chi_{k,x},\chi_{k,y},{r}_{k})$ is the mapping in \eqref{eqq32n}. The second-order Taylor series expansion at $\boldsymbol{\mu}_{\xi\rightarrow\mathbf{p}}$ is exploited to obtain the covariance
\begin{align}\label{eq35}
    \boldsymbol{C}_{\xi\rightarrow\mathbf{p}}=-\left.\frac{\partial^2\ln\Delta_{\xi\rightarrow\mathbf{p}}(\mathbf{p})}{\partial\mathbf{p}\partial\mathbf{p}^{\mathrm{T}}}\right|_{\mathbf{p}=\boldsymbol{\mu}_{\xi\rightarrow\mathbf{p}}}.
\end{align}
The detailed calculations of the gradient and the Hessian matrix in \eqref{eqq32} and \eqref{eq35} are provided in Appendices \ref{append1} and \ref{append2}, respectively. Algorithm \ref{GDM_algorithm} summarizes the steps for calculating $\boldsymbol{\mu}_{\xi\rightarrow\mathbf{p}}$ and $\boldsymbol{C}_{\xi\rightarrow\mathbf{p}}$.
\begin{algorithm}[t]
	\caption{Calculation of $\boldsymbol{\mu}_{\xi\rightarrow\mathbf{p}}$ and $\boldsymbol{C}_{\xi\rightarrow\mathbf{p}}$} 
	\label{GDM_algorithm} 
	{\bf Input:} Received signal $\boldsymbol{y}$, noise covariance matrix $\boldsymbol{C}_{n}$, beamforming matrix $\mathbf{W}$, and initializations $\boldsymbol{\chi}_{k}^{(0)}$ and $\boldsymbol{r}_{k}^{(0)}$.
 
	\begin{algorithmic}[1] 
        \REPEAT
        \STATE{Calculate the gradient with respect to $\boldsymbol{\chi}$ and $\boldsymbol{r}$ by \eqref{eqq71}.}
        \STATE{Obtain the step size $\iota_{{\chi}}^{(n)}$ and $\iota_{{r}}^{(n)}$ by Armijo
backtracking line search.}
        \STATE{Calculate $\boldsymbol{\chi}^{(n+1)}$ and $\boldsymbol{r}^{(n+1))}$ by \eqref{eqq32}.}
    \UNTIL $\|\boldsymbol{r}^{(n+1)} - \boldsymbol{r}^{(n)}\|<\epsilon$.
    \STATE{Calculate $\boldsymbol{\mu}_{\xi\rightarrow\mathbf{p}}$ and $\boldsymbol{C}_{\xi\rightarrow\mathbf{p}}$ by \eqref{eqq35n} and \eqref{eq35}, respectively.}
	\end{algorithmic}
 {\bf Output:} $\boldsymbol{\mu}_{\xi\rightarrow\mathbf{p}}$ and $\boldsymbol{C}_{\xi\rightarrow\mathbf{p}}$.
\end{algorithm}
\par
The message passing from $\mathbf{p}$ to $\xi$ is calculated by 
\begin{align}
    \Delta_{\mathbf{p}\rightarrow\xi}(\mathbf{p})\propto\Delta_{\psi\rightarrow\mathbf{p}}(\mathbf{p})p(\mathbf{p}),\label{eq30}
\end{align}
where $\Delta_{\psi\rightarrow\mathbf{p}}(\mathbf{p})$ is a Gaussian message calculated in \eqref{eqq43}. We further calculate \eqref{eq30} by
\begin{align}\label{eqq35}
    \Delta_{\mathbf{p}\rightarrow\xi}(\mathbf{p})=\mathcal{N}(\mathbf{p};\boldsymbol{\mu}_{\mathbf{p}\rightarrow\xi},\boldsymbol{C}_{\mathbf{p}\rightarrow\xi})
\end{align}
with mean $\boldsymbol{\mu}_{\mathbf{p}\rightarrow\xi}$ and covariance $\boldsymbol{C}_{\mathbf{p}\rightarrow\xi}$ given by
\begin{subequations}\label{eq38}
    \begin{align}
    \boldsymbol{C}_{\mathbf{p}\rightarrow\xi}^{-1}&=\boldsymbol{C}_{\psi\rightarrow\mathbf{p}}^{-1}+(\boldsymbol{C}_{\mathbf{p}}^\mathrm{pri})^{-1},\\
    \boldsymbol{\mu}_{\mathbf{p}\rightarrow\xi}&=\boldsymbol{C}_{\mathbf{p}\rightarrow\xi}\left(
    \boldsymbol{C}_{\psi\rightarrow\mathbf{p}}^{-1}\boldsymbol{\mu}_{\psi\rightarrow\mathbf{p}} + 
    {(\boldsymbol{C}_\mathbf{p}^\mathrm{pri})}^{-1}\boldsymbol{\mu}_{\mathbf{p}}^{\mathrm{pri}}\right),
\end{align}
\end{subequations}
where $\boldsymbol{C}_{\mathbf{p}}^{\mathrm{pri}}=\mathrm{blkdiag}\left(\nu_{1}^{\mathrm{pri}}\mathbf{I}_{3},...,\nu_{K}^{\mathrm{pri}}\mathbf{I}_{3}\right)$ and $\boldsymbol{\mu}_{\mathbf{p}}^{\mathrm{pri}}\allowbreak=\allowbreak\left[\allowbreak(\boldsymbol{\mu}_{\mathbf{p}_1}^{\mathrm{pri}})^{\mathrm{T}}\allowbreak,...,\allowbreak(\boldsymbol{\mu}_{\mathbf{p}_K}^{\mathrm{pri}})^{\mathrm{T}}\right]^{\mathrm{T}}$.
\subsubsection{Messages between $\xi$ and $\boldsymbol{\rho}$}The message passing from $\xi$ to $\boldsymbol{\rho}$ is calculated by 
\begin{align}\label{eq37}
    \Delta_{\xi\rightarrow\boldsymbol{\rho}}(\boldsymbol{\rho})\propto&\int_{\mathbf{p}}p\left(\boldsymbol{y}|\mathbf{p},\boldsymbol{\rho } \right)\Delta_{\mathbf{p}\rightarrow\xi}(\mathbf{p}).
\end{align}
The integral in \eqref{eq37} is difficult to calculate exactly due to the nonlinear relationship between $\mathbf{p}$ and $\mathbf{B}$. To simplify the integration with respect to {$\mathbf{p}$} in \eqref{eq37}, we introduce the approximation $\Delta_{\mathbf{p}\rightarrow\xi}=\delta(\mathbf{p}-\boldsymbol{\mu}_{\mathbf{p}\rightarrow\xi})$. This yields
\begin{align}\label{eq40}
    \Delta_{\xi\rightarrow\boldsymbol{\rho}}(\boldsymbol{\rho})\propto\mathcal{CN}\left(\boldsymbol{\rho};\boldsymbol{\mu}_{\xi\rightarrow\boldsymbol{\rho}},\boldsymbol{C}_{\xi\rightarrow\boldsymbol{\rho}}\right),
\end{align}
where
\begin{subequations}
\label{eq37n}
    \begin{align}
    \boldsymbol{C}_{\xi\rightarrow\boldsymbol{\rho}}&=\left(\hat{\mathbf{B}}^{\mathrm{H}}\boldsymbol{C}_{n}^{-1}\hat{\mathbf{B}}\right)^{-1},\label{eq37a_n}\\
    \boldsymbol{\mu}_{\xi\rightarrow\boldsymbol{\rho}}&=\boldsymbol{C}_{\xi\rightarrow\boldsymbol{\rho}}\hat{\mathbf{B}}^{\mathrm{H}}\boldsymbol{C}_{n}^{-1}\boldsymbol{y},\label{eq37b_n}
\end{align}
\end{subequations}
with $\hat{\mathbf{B}} = \left[\hat{\mathbf{B}}_{1},...,\hat{\mathbf{B}}_{K}\right]$. $\hat{\mathbf{B}}_{k}$ is obtained by substituting $\left[\boldsymbol{\mu}_{\mathbf{p}\rightarrow\xi}\right]_{3(k-1)+1:3k}$ into $\mathbf{B}_{k}$ in \eqref{eq21b} for $1\leq k\leq K$.
\par
The message from $\boldsymbol{\rho}$ to $\xi$ is equal to $\Delta_{\psi\rightarrow\boldsymbol{\rho}}(\boldsymbol{\rho})$ as calculated in \eqref{eq47}. We have
    \begin{align}\label{eq34}
    \Delta_{\boldsymbol{\rho}\rightarrow\xi}(\boldsymbol{\rho})\propto\mathcal{CN}\left(\boldsymbol{\rho};\boldsymbol{\mu}_{\boldsymbol{\rho}\rightarrow\xi},\boldsymbol{C}_{\boldsymbol{\rho}\rightarrow\xi}\right),
\end{align}
where $\boldsymbol{\mu}_{\boldsymbol{\rho}\rightarrow\xi}=\mathbf{0}$ and $\boldsymbol{C}_{\boldsymbol{\rho}\rightarrow\xi}=\boldsymbol{C}_{\psi\rightarrow\boldsymbol{\rho}}$.
\subsection{Calculation of Messages at Node $\psi$}\label{III-B}
\subsubsection{Messages between $\psi$ and $\mathbf{p}$}The message passing from $\psi$ to $\mathbf{p}$ is calculated by
\begin{align}\label{eq44}
    \Delta_{\psi\rightarrow\mathbf{p}}(\mathbf{p})\propto\int_{\boldsymbol{\rho},\boldsymbol{\varrho}}p\left( \boldsymbol{\rho }|\mathbf{p},\boldsymbol{\varrho}\right) \Delta_{\boldsymbol{\rho}\rightarrow\psi}(\boldsymbol{\rho})p\left( \boldsymbol{\varrho}\right),
\end{align}
where {$p\left( \boldsymbol{\rho }|\mathbf{p},\boldsymbol{\varrho}\right)$ is the geometric constrain in \eqref{7_17_eq19} and} $\Delta_{\boldsymbol{\rho}\rightarrow\psi}(\boldsymbol{\rho})$ is given by \eqref{eqq48}. Solving the integral in \eqref{eq44}, we obtain
\begin{align}\label{eqq42}
    \Delta_{\psi\rightarrow\mathbf{p}}(\mathbf{p})\propto\exp\left(\boldsymbol{\varXi}_{\psi}^{\mathrm{H}}\boldsymbol{\varUpsilon}_{\psi}^{-1}\boldsymbol{\varXi}_{\psi}\right)
\end{align}
where $\boldsymbol{\varXi}_{\psi} = \mathbf{C}^{\mathrm{H}}\boldsymbol{C}_{\boldsymbol{\rho}\rightarrow\psi}^{-1}\boldsymbol{\mu}_{\boldsymbol{\rho}\rightarrow\psi}$ and  $\boldsymbol{\varUpsilon}_{\psi} = (\boldsymbol{C}_{\boldsymbol{\varrho}}^{\mathrm{pri}})^{-1}+\mathbf{C}^{\mathrm{H}}\boldsymbol{C}_{\boldsymbol{\rho}\rightarrow\psi}^{-1}\mathbf{C}$ with $\mathbf{C}=\mathrm{blkdiag}\left(\mathbf{c}_{1},...,\mathbf{c}_{K}\right)$ and $\boldsymbol{C}_{\boldsymbol{\varrho}}^{\mathrm{pri}}=\mathrm{diag}([\tau_{1}^{\mathrm{pri}},...,\tau_{K}^{\mathrm{pri}}]^{\mathrm{T}})$. We approximate $\Delta_{\psi\rightarrow\mathbf{p}}(\mathbf{p})$ by a Gaussian distribution
\begin{align}\label{eqq43}
    \Delta_{\psi\rightarrow\mathbf{p}}(\mathbf{p})\propto\mathcal{N}(\mathbf{p};\boldsymbol{\mu}_{\psi\rightarrow\mathbf{p}},\boldsymbol{C}_{\psi\rightarrow\mathbf{p}}),
\end{align}
with the mean and covariance obtained by following the steps in \eqref{eq32}-\eqref{eq35}. As illustrated in Fig. \ref{fig:4}(c), $\ln\Delta_{\psi\rightarrow\mathbf{p}}(\mathbf{p})$ exhibits multiple local maximum with nearly identical function values.
This is because the spacing between adjacent subarray reference antennas is $N_{\mathrm{S},x}d$ (or $N_{\mathrm{S},y}d$), which exceeds $\frac{\lambda}{2}$, making two different user positions $\mathbf{p}_{i}$ and $\mathbf{p}_{j}$ has a similar array response $\mathbf{c}(\mathbf{p}_{i})\approx\mathbf{c}(\mathbf{p}_{j})$.
Consequently, the GA used in steps \eqref{eq32}-\eqref{eq35} requires a precise starting point to ensure Algorithm 1 converges to the global maximum. We introduce the initialization of $\boldsymbol{\mu}_{\psi\rightarrow\mathbf{p}}$ in Section \ref{IV-D}.
\par
The message from $\mathbf{p}$ to $\psi$ is calculated by
\begin{subequations}\label{eqq46}
    \begin{align}
    {\Delta_{\mathbf{p}\rightarrow\psi}(\mathbf{p})} &{\propto \Delta_{\xi\rightarrow\mathbf{p}}(\mathbf{p})p(\mathbf{p})}\\
    &\propto \mathcal{N}\left(\mathbf{p};\boldsymbol{\mu}_{\mathbf{p}\rightarrow\psi},\boldsymbol{C}_{\mathbf{p}\rightarrow\psi}\right),\label{eq40b}
    \end{align}
\end{subequations}

with mean $\boldsymbol{\mu}_{\mathbf{p}\rightarrow\psi}$ and covariance $\boldsymbol{C}_{\mathbf{p}\rightarrow\psi}$ calculated by
\begin{subequations}\label{eq48}
    \begin{align}
    \boldsymbol{C}_{\mathbf{p}\rightarrow\psi}^{-1}&=\boldsymbol{C}_{\xi\rightarrow\mathbf{p}}^{-1}+(\boldsymbol{C}_{\mathbf{p}}^{\mathrm{pri}})^{-1},\\
    \boldsymbol{\mu}_{\mathbf{p}\rightarrow\psi}&=\boldsymbol{C}_{\mathbf{p}\rightarrow\psi}\left(
    \boldsymbol{C}_{\xi\rightarrow\mathbf{p}}^{-1}\boldsymbol{\mu}_{\xi\rightarrow\mathbf{p}} + 
    (\boldsymbol{C}_{\mathbf{p}}^{\mathrm{pri}})^{-1}\boldsymbol{\mu}_{\mathbf{p}}^{\mathrm{pri}}\right).
\end{align}
\end{subequations}
\subsubsection{Messages between $\psi$ and $\boldsymbol{\rho}$}
The message passing from $\psi$ to $\boldsymbol{\rho}$ is calculated by
\begin{align}\label{eq47}
    \Delta_{\psi\rightarrow\boldsymbol{\rho}}(\boldsymbol{\rho})\propto\int_{\backslash\boldsymbol{\rho}}p\left( \boldsymbol{\rho }|\mathbf{p},\boldsymbol{\varrho}  \right)p(\boldsymbol{\varrho})\Delta_{\mathbf{p}\rightarrow\psi}(\mathbf{p}).
\end{align}
Similarly to the calculation in \eqref{eq37}, we introduce the approximation $\Delta_{\mathbf{p}\rightarrow\psi}(\mathbf{p})=\delta(\mathbf{p}-\boldsymbol{\mu}_{\mathbf{p}\rightarrow\psi})$ to simplify the integral in \eqref{eq47}, yielding
\begin{align}\label{eqq49}
    \Delta_{\psi\rightarrow\boldsymbol{\rho}}(\boldsymbol{\rho})\propto\mathcal{CN}(\boldsymbol{\rho};\boldsymbol{\mu}_{\psi\rightarrow\boldsymbol{\rho}},\boldsymbol{C}_{\psi\rightarrow\boldsymbol{\rho}}),
\end{align}
where $\boldsymbol{\mu}_{\psi\rightarrow\boldsymbol{\rho}}=\mathbf{0}$ and $\boldsymbol{C}_{\psi\rightarrow\boldsymbol{\rho}}=\hat{\mathbf{C}}\boldsymbol{C}_{\boldsymbol{\varrho}}^{\mathrm{pri}}\hat{\mathbf{C}}^{\mathrm{H}}$. Here, $\hat{\mathbf{C}}=\mathrm{blkdiag}(\hat{\mathbf{c}}_{1},...,\hat{\mathbf{c}}_{K})$, and $\hat{\mathbf{c}}_{k}$ is obtained by substituting $\left[\boldsymbol{\mu}_{\mathbf{p}\rightarrow\psi}\right]_{3(k-1)+1:3k}$ into $\mathbf{c}_{k}(\mathbf{p}_{k})$ in \eqref{eq18} for $1\leq k\leq K$.
\par
The message from $\boldsymbol{\rho}$ to $\psi$ is equal to $\Delta_{\xi\rightarrow\boldsymbol{\rho}}(\boldsymbol{\rho})$ as calculated in \eqref{eq40}. We have
\begin{align}\label{eqq48}
    \Delta_{\boldsymbol{\rho}\rightarrow\psi}(\boldsymbol{\rho})\propto\mathcal{CN}\left(\boldsymbol{\rho};\boldsymbol{\mu}_{\boldsymbol{\rho}\rightarrow\psi},\boldsymbol{C}_{\boldsymbol{\rho}\rightarrow\psi}\right),
\end{align}
where $\boldsymbol{\mu}_{\boldsymbol{\rho}\rightarrow\psi}=\boldsymbol{\mu}_{\xi\rightarrow\boldsymbol{\rho}}$ and $\boldsymbol{C}_{\boldsymbol{\rho}\rightarrow\psi}=\boldsymbol{C}_{\xi\rightarrow\boldsymbol{\rho}}$.
\subsubsection{Messages between $\psi$ and $\boldsymbol{\varrho}$} The message passing from $\psi$ to $\boldsymbol{\varrho}$ is calculated by
\begin{align}
    \Delta_{\psi\rightarrow\boldsymbol{\varrho}}(\boldsymbol{\varrho})&\propto\int_{\backslash\boldsymbol{\varrho}}p\left( \boldsymbol{\rho }|\mathbf{p},\boldsymbol{\varrho}\right) \Delta_{\boldsymbol{\rho}\rightarrow\psi}(\boldsymbol{\rho}){\Delta_{\mathbf{p}\rightarrow\psi}(\mathbf{p})}\notag\\
    &\propto\mathcal{CN}\left(\boldsymbol{\varrho};\boldsymbol{\mu}_{\psi\rightarrow\boldsymbol{\varrho}},\boldsymbol{C}_{\psi\rightarrow\boldsymbol{\varrho}}\right),
\end{align}
where 
\begin{subequations}
    \begin{align}
    \boldsymbol{C}_{\psi\rightarrow\boldsymbol{\varrho}}&=\left(\hat{\mathbf{C}}^{\mathrm{H}}\boldsymbol{C}_{\boldsymbol{\rho}\rightarrow\psi}^{-1}\hat{\mathbf{C}}\right)^{-1},\\
    \boldsymbol{\mu}_{\psi\rightarrow\boldsymbol{\varrho}}&=\boldsymbol{C}_{\psi\rightarrow\boldsymbol{\varrho}}\hat{\mathbf{C}}^{\mathrm{H}}\boldsymbol{C}_{\boldsymbol{\rho}\rightarrow\psi}^{-1}\boldsymbol{\mu}_{\boldsymbol{\rho}\rightarrow\psi},
\end{align}
\end{subequations}
\par
The message from $\boldsymbol{\varrho}$ to $\psi$ is equal to the prior distribution $p(\boldsymbol{\varrho})$. 
\subsection{Output Estimates}
The marginal message at $\mathbf{p}$ is calculated by
\begin{subequations}\label{eqq54}
    \begin{align}
    \Delta_{\mathbf{p}}(\mathbf{p})&\propto\Delta_{\xi\rightarrow\mathbf{p}}(\mathbf{p})\Delta_{\psi\rightarrow\mathbf{p}}(\mathbf{p})p(\mathbf{p})\\
    &\propto\mathcal{N}(\mathbf{p};\boldsymbol{\mu}_{\mathbf{p}},\boldsymbol{C}_{\mathbf{p}}),
\end{align}
\end{subequations}
with mean $\boldsymbol{\mu}_{\mathbf{p}}$ and covariance $\boldsymbol{C}_{\mathbf{p}}$ given by
\begin{subequations}
    \begin{align}
    \boldsymbol{C}_{\mathbf{p}}^{-1} &= \boldsymbol{C}_{\xi\rightarrow\mathbf{p}}^{-1}+\boldsymbol{C}_{\psi\rightarrow\mathbf{p}}^{-1}+(\boldsymbol{C}_{\mathbf{p}}^{\mathrm{pri}})^{-1},\\
    \boldsymbol{\mu}_{\mathbf{p}}&=\boldsymbol{C}_{\mathbf{p}}\left(\boldsymbol{C}_{\xi\rightarrow\mathbf{p}}^{-1}\boldsymbol{\mu}_{\xi\rightarrow\mathbf{p}}+
    \boldsymbol{C}_{\psi\rightarrow\mathbf{p}}^{-1}\boldsymbol{\mu}_{\psi\rightarrow\mathbf{p}} + 
    (\boldsymbol{C}_{\mathbf{p}}^{\mathrm{pri}})^{-1}\boldsymbol{\mu}_{\mathbf{p}}^{\mathrm{pri}}\right).
\end{align}
\end{subequations}
The output estimation of user position $\mathbf{p}_{k}$ is given by $\hat{\mathbf{p}}_{k} = \left[\boldsymbol{\mu}_{\mathbf{p}}\right]_{3(k-1)+1:3k}$. 
\par
The marginal message at $\boldsymbol{\varrho}$ is calculated by
\begin{subequations}\label{eqq52}
    \begin{align}
    \Delta_{\boldsymbol{\varrho}}(\boldsymbol{\varrho})&\propto\Delta_{\psi\rightarrow\boldsymbol{\varrho}}(\boldsymbol{\varrho}) p\left( \boldsymbol{\varrho}\right)\\
    &\propto\mathcal{CN}( \boldsymbol{\varrho};\boldsymbol{\mu}_{\boldsymbol{\varrho}},\boldsymbol{C}_{\boldsymbol{\varrho}})
\end{align}
\end{subequations}
with the mean and covariance given by
\begin{subequations}
    \begin{align}
    \boldsymbol{C}_{\boldsymbol{\varrho}}&=\left(\boldsymbol{C}_{\psi\rightarrow\boldsymbol{\varrho}}^{-1}+\left(\boldsymbol{C}_{\boldsymbol{\varrho}}^{\mathrm{pri}}\right)^{-1}\right)^{-1}\\
    \boldsymbol{\mu}_{\boldsymbol{\varrho}}&=\boldsymbol{C}_{\boldsymbol{\varrho}}\boldsymbol{C}_{\psi\rightarrow\boldsymbol{\varrho}}^{-1}\boldsymbol{\mu}_{\psi\rightarrow\boldsymbol{\varrho}}
\end{align}
\end{subequations}
{The output estimation of reference array gain $\varrho_{k}$ is given by $\hat{\varrho}_{k} = \left[\boldsymbol{\mu}_{\boldsymbol{\varrho}}\right]_{k}$.}
\par
Based on the estimated user positions $\boldsymbol{\mu}_{\mathbf{p}}$ and the reference array gain $\boldsymbol{\mu}_{\boldsymbol{\varrho}}$, we obtain $\hat{h}_{ref,k}^{(\frac{M}{2},\frac{M}{2})}=\frac{[\boldsymbol{\mu}_{\boldsymbol{\varrho}}]_{k}}{\sqrt{P_{k}}x_{k}}$ and reconstruct the channel $\hat{\boldsymbol{H}}_{k}$ by \eqref{eq20}\footnote{When $x_{k}$ is unknown, we cannot estimate $h_{ref,k}^{(\frac{M}{2},\frac{M}{2})}$ from $\boldsymbol{\mu}_{\boldsymbol{\varrho}}$. In this case, we reconstruct the equivalent channel $\sqrt{P_{k}}x_{k}\boldsymbol{H}_{k}$.
}.
\subsection{Initialization} \label{IV-D}
We initialize user positions $\{\mathbf{p}_{k}\}_{k=1}^{K}$ sequentially. In the $t$-th step, $\mathbf{p}_{k}$ is fix to its initialization $\mathbf{p}_{k}=\mathbf{p}_{k}^{\mathrm{ini}}$ for $1 \leq k \leq t-1$.
Then, we initialize the $t$-th user's position $\mathbf{p}_{t}$ by following the steps in \eqref{eq31}-\eqref{eq35}, where \eqref{eq31} is maximized with respect to $\mathbf{p}_{t}${, and $\boldsymbol{C}_{\boldsymbol{\rho}\rightarrow\xi}^{-1}$ is initialized as $\mathbf{0}$}. To obtain the starting point of the GA iterations in \eqref{eqq32}, we represent the user position $\mathbf{p}_{t}$ in the polar domain as $(\chi_{t,x}, \chi_{t,y}, r_{t})$ and construct a 3D discrete grid as
\begin{align}\label{eq56}
    &\mathcal{G}\overset{\Delta}{=}\left\{(\chi_{x}^{i}, \chi_{y}^{j}, r^{l}) \mid \chi_{x}^{i} = -1 + \frac{2(i-1)}{M_{x}}, i \in \{1, \ldots, M_{x}\}, \right.\notag\\
    &\quad\left.\chi_{y}^{j} = -1 + \frac{2(j-1)}{M_{y}}, j \in \{1, \ldots, M_{y}\}, \right.\notag\\
    &\quad\left.r^{l} = r_{\mathrm{min}} + \frac{r_{\mathrm{max}}-r_{\mathrm{min}}}{M_{r}-1}(l-1), l \in \{1, \ldots, M_{r}\}\right\},
\end{align}
where $M_{x}$, $M_{y}$, and $M_{r}$ are the numbers of grid points for $\chi_{x}$, $\chi_{y}$, and $r$, respectively. The parameters $r_{\mathrm{max}}$ and $r_{\mathrm{min}}$ define the range of the area of interest. We perform an exhaustive search over the discrete grid $\mathcal{G}$ to find the grid point $(\chi_{x}^{*}, \chi_{y}^{*}, r^{*}) \in \mathcal{G}$ that maximizes \eqref{eq31}, which serves as the starting point for the GA iteration in \eqref{eqq32}. The position initialization $\mathbf{p}_{t}^{\mathrm{ini}}$ is obtained when the GA iteration converges. 
After all $K$ users have been initialized, we initialize $\boldsymbol{\mu}_{\xi\rightarrow\mathbf{p}}=[(\mathbf{p}_{1}^{\mathrm{ini}})^{\mathrm{T}}, \ldots, (\mathbf{p}_{K}^{\mathrm{ini}})^{\mathrm{T}}]^{\mathrm{T}}$, $\boldsymbol{\mu}_{\mathbf{p}\rightarrow\xi}=\boldsymbol{\mu}_{\xi\rightarrow\mathbf{p}}$, and $\boldsymbol{\mu}_{\psi\rightarrow\mathbf{p}}=\boldsymbol{\mu}_{\xi\rightarrow\mathbf{p}}$. The covariance matrices $\boldsymbol{C}_{\xi\rightarrow\mathbf{p}}$ and $\boldsymbol{C}_{\mathbf{p}\rightarrow\xi}$ are initialized as $\sigma_{\mathrm{ini}}^{2}\mathbf{I}$, where $\sigma_{\mathrm{ini}}^{2}$ is a large positive value.
\begin{algorithm}[t]
	\caption{APLE-LM} 
	\label{APLE-LM_algorithm} 
	{\bf Input:} Received signal $\boldsymbol{y}$, noise power $\sigma_{n}^{2}$, power radiation pattern $F(\theta,\phi)$, beamforming matrix $\mathbf{W}$, and prior distributions $p(\varrho_{k})$ and $p(\mathbf{p}_{k})$ for $1\leq k\leq K$.
 
	\begin{algorithmic}[1] 
        \STATE{\bf Initialization:} $\boldsymbol{\mu}_{\xi\rightarrow\mathbf{p}}$, $\boldsymbol{\mu}_{\mathbf{p}\rightarrow\xi}$, $\boldsymbol{\mu}_{\psi\rightarrow\mathbf{p}}$, $\boldsymbol{C}_{\xi\rightarrow\mathbf{p}}$, and $\boldsymbol{C}_{\mathbf{p}\rightarrow\xi}$ are initialized by following the steps in Section \ref{IV-D}.
        \REPEAT
        \STATE{Update $\boldsymbol{\mu}_{\mathbf{p}\rightarrow\psi}$ and $\boldsymbol{C}_{\mathbf{p}\rightarrow\psi}$ by \eqref{eq48}.}
        \STATE{Update $\boldsymbol{\mu}_{\psi\rightarrow\boldsymbol{\rho}}=\mathbf{0}$ and $\boldsymbol{C}_{\psi\rightarrow\boldsymbol{\rho}}=\hat{\mathbf{C}}\boldsymbol{C}_{\boldsymbol{\varrho}}^{\mathrm{pri}}\hat{\mathbf{C}}^{\mathrm{H}}$.}
        \STATE{Update $\boldsymbol{\mu}_{\boldsymbol{\rho}\rightarrow\xi}=\mathbf{0}$ and $\boldsymbol{C}_{\boldsymbol{\rho}\rightarrow\xi}=\boldsymbol{C}_{\psi\rightarrow\boldsymbol{\rho}}$.}
        \STATE{Update $\boldsymbol{\mu}_{\xi\rightarrow\mathbf{p}}$ and $\boldsymbol{C}_{\xi\rightarrow\mathbf{p}}$ by \textbf{Algorithm \ref{GDM_algorithm}}}
        \STATE{Update $\boldsymbol{\mu}_{\xi\rightarrow\boldsymbol{\rho}}$ and $\boldsymbol{C}_{\xi\rightarrow\boldsymbol{\rho}}$ by \eqref{eq37n}.}
        \STATE{Update $\boldsymbol{\mu}_{\boldsymbol{\rho}\rightarrow\psi}=\boldsymbol{\mu}_{\xi\rightarrow\boldsymbol{\rho}}$ and $\boldsymbol{C}_{\boldsymbol{\rho}\rightarrow\psi}=\boldsymbol{C}_{\xi\rightarrow\boldsymbol{\rho}}$.}
        \STATE{Update $\boldsymbol{\mu}_{\psi\rightarrow\mathbf{p}}$ and $\boldsymbol{C}_{\psi\rightarrow\mathbf{p}}$ by \textbf{Algorithm \ref{GDM_algorithm}}.}
        \STATE{Update $\boldsymbol{\mu}_{\mathbf{p}\rightarrow\xi}$ and $\boldsymbol{C}_{\mathbf{p}\rightarrow\xi}$ by \eqref{eq38}.}
    \UNTIL stopping criterion is met.
    \STATE{Calculate the marginal message $\Delta_{\mathbf{p}}(\mathbf{p})$ and $\Delta_{\boldsymbol{\varrho}}(\boldsymbol{\varrho})$ by \eqref{eqq54} and \eqref{eqq52}, respectively.}
	\end{algorithmic}
 {\bf Output:} Estimations of user positions  $\hat{\mathbf{p}}_{k}$ and channel reconstructions $\hat{\boldsymbol{H}}_{k}$ for $1\leq k \leq K$. Estimation of reference
array gain vector $\hat{\boldsymbol{\varrho}}=\boldsymbol{\mu}_{\boldsymbol{\varrho}}$.
\end{algorithm}
\subsection{Overall Algorithm}
We summarize the proposed APLE-LM algorithm in {Algorithm \ref{APLE-LM_algorithm}}. The initialization of messages follows the steps in Section \ref{IV-D}. The message calculations are generally divided into two parts: lines 3 to 6 correspond to the {clockwise} message updating, and lines 7 to 10 correspond to the {counterclockwise} message updating. The stopping criterion is defined as the norm of the relative change in $\boldsymbol{\mu}_{\mathbf{p}}$ between two adjacent iterations being smaller than the predetermined parameter $\epsilon$ for each $1\leq k\leq K$. 
{Note that in the factor graph in Fig. \ref{factor}, the variable nodes $\mathbf{p}$ and $\boldsymbol{\rho}$, along with the check nodes $\psi$ and $\xi$, form a cycle, making the factor graph loopy. Due to the presence of this cycle, the iterative message-passing process may suffer from instability or slow convergence; therefore, we apply the damping technique to the messages $\Delta_{\mathbf{p} \rightarrow \xi}(\mathbf{p})$ and $\Delta_{\xi \rightarrow \mathbf{p}}(\mathbf{p})$ to improve convergence\cite{parker2014bilinear}.}
\par
The complexity of the proposed APLE-LM algorithm primarily arises from the exhaustive search on the 3D discrete grid \eqref{eq56} to initialize $\boldsymbol{\mu}_{\xi \rightarrow \mathbf{p}}$ and the calculation of messages $\Delta_{\xi \rightarrow \mathbf{p}}(\mathbf{p})$.
{Since the grid points are fixed, we can compute $\boldsymbol{C}_{n}^{-1}\mathbf{B}\left(\mathbf{B}^{\mathrm{H}}\boldsymbol{C}_{n}^{-1}\mathbf{B}\right)^{-1}\mathbf{B}\boldsymbol{C}_{n}^{-1}$ in \eqref{eq31} for each grid point in advance to decrease complexity of the initialization. Given that the number of grid points in the three-dimensional search grid is \(M_{x} M_{y} M_{r}\), and \(K\) users need initialization, the complexity of the grid search initialization is \(O(KM_{x} M_{y} M_{r}N_{RF})\). The complexity of computing message $\Delta_{\xi \rightarrow \mathbf{p}}(\mathbf{p})$ depends on the gradient computation and the Armijo backtracking line search. The complexity of the gradient computation is dominated by \eqref{eq655}, where the computation of \(\partial\mathbf{B}\) has a complexity of \(O(KN_{\mathrm{B}}N_\mathrm{RF})\), thereby, the gradient computation has a complexity of \(O(KM^2N_{\mathrm{RF}}^2+KN_{\mathrm{B}}N_{\mathrm{RF}})\). The Armijo backtracking line search requires multiple evaluations of the objective function to select an appropriate step size, with a complexity of \(O(KN_{\mathrm{B}}N_{\mathrm{RF}})\) for each evaluation. Therefore, the total complexity of the proposed APLE-LM algorithm is \(\mathcal{O}(KM_{x} M_{y} M_{r}N_{RF}+n_{1}(KM^2N_{\mathrm{RF}}^2+n_{2}KN_{\mathrm{B}}N_{\mathrm{RF}}))\), where $n_{1}$ is the number of iterations in GA and $n_{2}$ is the average number of evaluating objective function in the Armijo backtracking line search. Generally, the number of ELAA antenna is large and $n_2N_{\mathrm{B}}\gg M^2N_{\mathrm{RF}}$, we obtain the complexity as $\mathcal{O}(KM_{x} M_{y} M_{r}N_{RF}+n_{1}n_{2}KN_{\mathrm{B}}N_{\mathrm{RF}})$.
}\par
\section{Bayesian Cram\'er-Rao Bound}
For the Bayesian inference problem in Section \ref{III-C}, the BCRB is a commonly used lower bound on the MSE of parameter estimation\cite{van2004detection}. In this section, we derive the BCRB to benchmark the user localization performance and channel reconstruction performance. Given the received signal model \eqref{eq9}, we define parameter set $\boldsymbol{\eta}=\left[\mathbf{p}_{1}^{\mathrm{T}},\Re\{\varrho_{1}\},\Im\{\varrho_{1}\},...,\mathbf{p}_{K}^{\mathrm{T}},\Re\{\varrho_{K}\},\Im\{\varrho_{K}\}\right]^{\mathrm{T}}\in\mathbb{R}^{5K}$. The BCRB is derived from the information matrix $\mathbf{J}$ calculated by
\begin{align}
    \mathbf{J} = \mathbf{J}_{\mathrm{F}} + \mathbf{J}_{\mathrm{P}}
\end{align}
where $\mathbf{J}_{\mathrm{F}}$ and $\mathbf{J}_{\mathrm{P}}$ are the Fisher information matrix and the \textit{a priori} information matrix, respectively. The matrices of $\mathbf{J}_{\mathrm{F}}$ and $\mathbf{J}_{\mathrm{P}}$ are calculated by
\begin{align}
    [\mathbf{J}_{\mathrm{F}}]_{i,j} = \mathbb{E}\left[-\frac{\partial^2\ln \left(p\left( \boldsymbol{y}|\mathbf{p}, \boldsymbol{\rho }\right) p\left( \boldsymbol{\rho }|\mathbf{p},\boldsymbol{\varrho}\right)\right)}{\partial\eta_{i}\partial\eta_{j}}\right],\label{eq50}\\
    [\mathbf{J}_{\mathrm{P}}]_{i,j} = \mathbb{E}\left[-\frac{\partial^2\ln
    \left(p\left( \boldsymbol{\varrho} \right)\prod_{k=1}^K{  p\left( \mathbf{p}_k \right)}\right)}{\partial\eta_{i}\partial\eta_{j}}\right].\label{eq51}
\end{align}
By plugging \eqref{eq24}, \eqref{7_17_eq19} and \eqref{eqq25} into \eqref{eq50} and \eqref{eq51}, we obtain
\begin{align}
    \mathbf{J}_{\mathrm{F}} &= 2\Re\left\{\left(\frac{\partial(\mathbf{B}\boldsymbol{\rho})}{\partial\boldsymbol{\eta}}\right)^{\mathrm{H}}
    \boldsymbol{C}_{n}^{-1}
    \left(\frac{\partial(\mathbf{B}\boldsymbol{\rho})}{\partial\boldsymbol{\eta}}\right)\right\},\\
    \mathbf{J}_{\mathrm{P}} &= \mathrm{blkdiag}\left(\nu_{1}^{\mathrm{pri}}\mathbf{I}_{3},\frac{\tau_{1}^{\mathrm{pri}}}{2}\mathbf{I}_{2},...,
    \nu_{K}^{\mathrm{pri}}\mathbf{I}_{3},\frac{\tau_{K}^{\mathrm{pri}}}{2}\mathbf{I}_{2}\right).
\end{align}
The BCRB of the parameter estimation $\hat{\boldsymbol{\eta}}=[\hat{\mathbf{p}}_{1}^{\mathrm{T}}, \allowbreak \Re\{\hat{\varrho}_{1}\}, \allowbreak \Im\{\hat{\varrho}_{1}\}, \allowbreak ..., \allowbreak \hat{\mathbf{p}}_{K}^{\mathrm{T}}, \allowbreak \Re\{\hat{\varrho}_{K}\}, \allowbreak \Im\{\hat{\varrho}_{K}\}]^{\mathrm{T}}$ is given by
\begin{align}
    \mathbb{E}\left[\left(\left[\boldsymbol{\hat{\eta}}\right]_{i}-\left[\boldsymbol{\eta}\right]_{i}\right)^2\right]\geq\left[\mathbf{J}^{-1}\right]_{i,i}.
\end{align}
\par
The near-field channel $\boldsymbol{H}_{k}$ is parameterized by $\boldsymbol{\eta}$ as shown in \eqref{eq20}. We construct the real-valued channel vector $\boldsymbol{h}=\left[\Re\{\boldsymbol{h}_{1}^{\mathrm{T}}\},\Im\{\boldsymbol{h}_{1}^{\mathrm{T}}\},...,\Re\{\boldsymbol{h}_{K}^{\mathrm{T}}\},\Im\{\boldsymbol{h}_{K}^{\mathrm{T}}\}\right]^{\mathrm{T}}\in \mathbb{C}^{2KN_{\mathrm{B}}}$. 
The BCRB of the real-valued channel vector channel estimation $\hat{\boldsymbol{h}}$ is obtained by exploiting parameter transformation as
\begin{align}
    \mathbb{E}\left[\left([\boldsymbol{\hat{h}}]_{i}-\left[\boldsymbol{h}\right]_{i}\right)^2\right]\geq\left[\frac{\partial\boldsymbol{h}}{\partial\boldsymbol{\eta}}\mathbf{J}^{-1}\left(\frac{\partial\boldsymbol{h}}{\partial\boldsymbol{\eta}}\right)^{\mathrm{T}}\right]_{i,i}.
\end{align}
\begin{table}[t]
    \centering
    \caption{Default System Parameters}
    \label{SysParaTable}
    \begin{tblr}{
        width = 0.98\linewidth, 
        colspec = {|
           Q[3.5,l,m]
           Q[2,c,m]
           Q[2,c,m]
           |
        }, 
        rowhead = 1,
        row{even} = {gray9},
        cell{1}{3} = {c=1}{c}, 
      }
      \toprule[1.5pt]
        \textbf{Parameter} & \textbf{Symbol} & {{{\textbf{Value}}}} \\
      \midrule[1pt]
        Carrier frequency      & $f$      & \SI{6}{GHz}\\ 
        Wavelength       & $\lambda$      & \SI{0.05}{m}      \\ 
        Antenna interval & $d$ &\SI{0.025}{m}\\
        Number of BS antennas & $N_{\mathrm{B},x}\times N_{\mathrm{B},y}$ & $45\times45$ \\
        {Number of subarray antennas} & {${N_{\mathrm{S},x}}\times{N_{\mathrm{S},y}}$} & {$15\times15$}\\
        Number of RF chains & $N_{\mathrm{RF}}$ & $160$ \\
        Number of UEs & $K$ & $3$ \\
        BS-UE distance range     & $[r_{\mathrm{min}},r_{\mathrm{max}}]$      & {$[\SI{5}{m},\SI{10}{m}]$}         \\ 
      \bottomrule[1.5pt]
      \end{tblr}
\end{table}
\section{Numerical Results}\label{S6}
In this section, we conduct numerical experiments under various parameter configurations to evaluate the proposed APLE-LM algorithm. The performance of the proposed APLE-LM algorithm is compared with the Bayesian Cramer Rao bound, as well as other state-of-the-art user localization algorithms.
\subsection{Simulation Settings}
In the numerical experiments, the settings of the uplink multiuser MIMO systems are listed in Table \ref{SysParaTable}, unless otherwise specified. The UPA of the BS has equal size in both the $x$-axis and the $y$-axis. 
The Rayleigh distance $\frac{2D^{2}}{\lambda}$ is a widely used criterion to distinguish far-field region and near-field region, where $D=(N_{\mathrm{B},x}^2d^2+N_{\mathrm{B},y}^{2}d^2)^{\frac{1}{2}}$ is the aperture of the antenna array \cite{selvan2017fraunhofer}. The Rayleigh distance under the default parameter settings is $\SI{101.25}{m}$. The $K$ UE positions are generated randomly within a cone-shaped region in the near-field of the BS antenna array. The vertex angle of the cone region is $120$ degrees and the BS-UE distance range is bounded by $r_{\mathrm{min}}$ and $r_{\mathrm{max}}$. 
Each element of analog beamforming matrix $\mathbf{W}$ is unimodular with the phase randomly sampled from $[0,2\pi]$. 
{To demonstrate the robustness of the APLE-LM algorithm in the absence of prior knowledge about the user positions $\mathbf{p}_{k}$ and the reference array gain $\alpha$, we assign non-informative priors to $\mathbf{p}_{k}$ and $\varrho_{k}$ by setting $\tau_{k}^{\mathrm{pri}} = 10^9$ and $\nu_{k}^{\mathrm{pri}}=10^9$ for all $1 \leq k \leq K$.}
\par
The performance of the proposed APLE-LM algorithm is evaluated by the root-MSE (RMSE) of the estimates of user positions $\mathbf{p}_{k}$ and the normalized-MSE (NMSE) of the channel reconstruction, respectively defined as
\begin{subequations}\label{eq61}
\begin{align}
\label{RMSE}
    \mathrm{RMSE}(\mathbf{p}_{k})=\sqrt{\frac{1}{K}\mathbb{E}\left[\sum_{k=1}^{K}\left\|\hat{\mathbf{p}}_{k}-\mathbf{p}_{k}\right\|_{2}^{2}\right]},\\
    \mathrm{NMSE}(\boldsymbol{h}_{k})=\frac{1}{K}\mathbb{E}\left[\sum_{k=1}^{K}\frac{\left\|\hat{\boldsymbol{h}}_{k}-\boldsymbol{h}_{k}\right\|_{2}^{2}}{\left\|\boldsymbol{h}_{k}\right\|_{2}^{2}}\right].
\end{align}
\end{subequations}
The expectations in \eqref{eq61} are numerically approximated by averaging over 1000 random experiments.
\par
We consider the following two baseline schemes:
\begin{enumerate}
    \item \textbf{ES-GA}: The user positions are estimated sequentially. For $1 \leq k \leq K$, the position estimate $\hat{\mathbf{p}}_{k}$ is obtained by first performing an exhaustive search (ES) over the grid $\mathcal{G}$ to find the on-grid maximum of the objective function
    \begin{align}
        -\left(\boldsymbol{y}_{\mathrm{res}}-\hat{\beta}_{k}\mathbf{W}\mathbf{a}(\mathbf{p}_{k})\right)^{\mathrm{H}}\boldsymbol{C}_{n}^{-1}\left(\boldsymbol{y}_{\mathrm{res}}-\hat{\beta}_{k}\mathbf{W}\mathbf{a}(\mathbf{p}_{k})\right),\label{eqq64}
    \end{align}
    where $\hat{\beta}_{k}=\frac{\mathbf{a}(\mathbf{p}_{k})^{\mathrm{H}}\mathbf{W}^{\mathrm{H}}\boldsymbol{C}_{n}^{-1}\boldsymbol{y}_{\mathrm{res}}}{(\mathbf{a}(\mathbf{p}_{k})^{\mathrm{H}}\mathbf{W}^{\mathrm{H}}\boldsymbol{C}_{n}^{-1}\mathbf{W}\mathbf{a}(\mathbf{p}_{k}))}$. This is followed by applying the GA method to find an off-grid local maximum. The residual signal $\boldsymbol{y}_{\mathrm{res}}$ is then updated by subtracting the signal component corresponding to user $k$.
    \item \textbf{Polar-Domain SIGW (P-SIGW) \cite{cui2022channel}}: This algorithm represents user locations in the polar domain as given in \eqref{eqq32n}, constructs a polar domain dictionary matrix, and uses the Simultaneous Orthogonal Matching Pursuit (SOMP) algorithm for initial estimation of $(\hat{\boldsymbol{\chi}},\hat{\boldsymbol{r}})$. It then refines the estimations using the classical off-grid Simultaneous Gridless Weighted (SIGW) algorithm.
\end{enumerate}
When the antenna power radiation pattern is unknown, a variant of the proposed APLE-LM can be employed, using the channel approximation in \eqref{eq11}. {This variant, referred to as APLE-LM with Approximate Channel Model (APLE-LM-ACM), ignores the directional characteristics of the antenna radiation pattern and the distance-dependent free-space path loss differences between antennas.}
\par
{Similar to the complexity of the APLE-LM algorithm, The complexities of the baseline algorithms ES-GA and P-SIGW are dominated by the initialization grid search and gradient descent. Based on the objective function \eqref{eqq64} and \cite{cui2022channel}, the search complexity is identical for both ES-GA and P-SIGW, with an initialization grid search complexity of \(\mathcal{O}(KM_{x} M_{y} M_{r}N_{\mathrm{RF}})\). For gradient descent, the complexity of calculating the gradient for both baseline algorithms ES-GA and P-SIGW is \(\mathcal{O}(N_{\mathrm{RF}}^2)\). The complexity of evaluating the objective function is \(\mathcal{O}(N_{\mathrm{B}}N_{\mathrm{RF}})\). Therefore, the overall complexity for ES-GA and P-SIGW is \(\mathcal{O}(KM_{x} M_{y} M_{r}N_{\mathrm{RF}} + n_{1}(KN_{\mathrm{RF}}^2 + n_{2}KN_{\mathrm{B}}N_{\mathrm{RF}}))\), where \(n_{1}\) is the number of iterations in the gradient ascent (GA) process, and \(n_{2}\) is the average number of searches in the Armijo backtracking line search. Given that \(n_2N_{\mathrm{B}} \gg M^2N_{\mathrm{RF}}\), the complexity can be simplified to \(\mathcal{O}(KM_{x} M_{y} M_{r}N_{\mathrm{RF}} + n_{1}n_{2}KN_{\mathrm{B}}N_{\mathrm{RF}})\). We note that the complexity of the APLE-LM algorithm and the baseline schemes share the same mathematical form. However, as demonstrated in the next subsection, the proposed APLE-LM algorithm significantly reduces the number of grid points required for initialization, thereby effectively lowering the initialization complexity.
}
\subsection{Results and Discussions}
\begin{figure}[t]
    \centering
    \resizebox{0.9\linewidth}{!}{\includegraphics{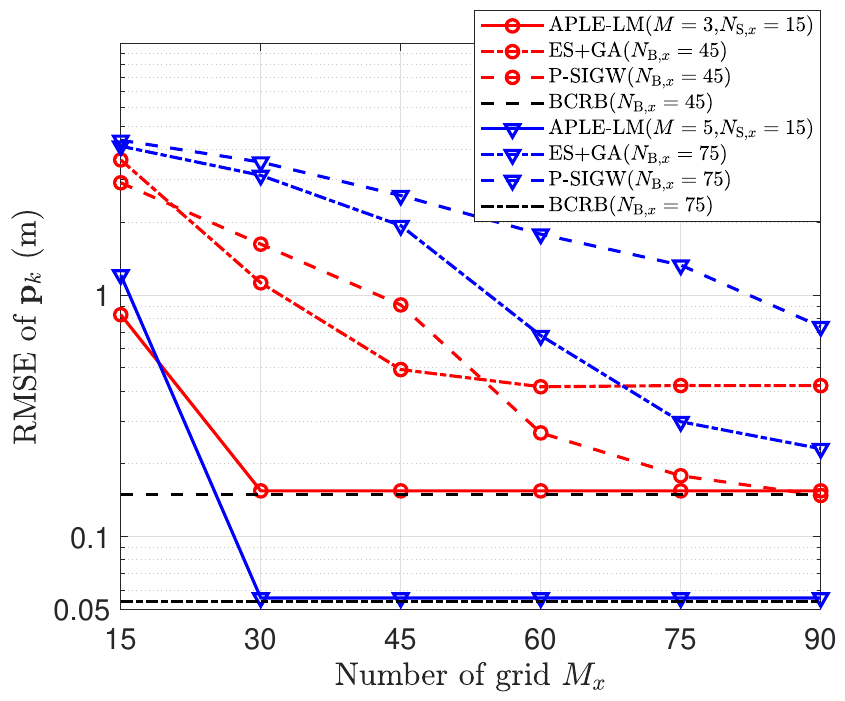}}
    \caption{The UE localization performance v.s. the number of grid points $M_{x}$. The SNR is fixed at $\SI{15}{dB}$.}
    \label{Fig_gridsize}
\end{figure}
\subsubsection{Number of grid points}
Fig. 4 illustrates the impact of the number of $\left\{\chi_{x}, \chi_{y}\right\}$ grid points on the performance of the localization algorithm. We set the number of grid points for $\chi_{x}$ and $\chi_{y}$ to be equal, i.e., $M_{x} = M_{y}$. The number of grid points in the range domain $r$ is fixed at $M_{r} = 2$. For comparison, we evaluate the performance of the baseline schemes ES+GA and P-SIGW. Both the proposed APLE-LM algorithm and baseline schemes have lower localization accuracy when the number of grid points is $M_{x} = 15$. This is because, with a small number of grid points, both algorithms tend to get trapped in local optima, leading to large localization errors. Increasing the number of grid points alleviates this issue. When the number of grid points reaches $M_{x} \geq 45$, the APLE-LM algorithm approaches the performance bound. In contrast, the ES+GA and P-SIGW algorithms require a higher number of grid points to achieve stable localization performance. Specifically, with 45 array antennas, the ES+GA algorithm requires $M_{x} = 60$ grid points, and with 75 array antennas, it requires $M_{x} > 90$ grid points. The number of grid points required by the baseline algorithms increase with array size, whereas the number of grid points required by the APLE-LM algorithm remains the same as the array size increases from $45$ to $75$. {For $N_{\mathrm{B},x}=45$, baseline algorithm P-SIGW requires tripling the number of angular grid points in one angular domain to achieve the same localization performance, resulting in a total number of initialization grid points that is 9 times (two angular domain) that of the APLE-LM algorithm. When the array size increases to 75, this effect becomes even more pronounced, further validating that APLE-LM can effectively reduce the number of initialization grid points in the ELAA near-field localization problem, thereby lowering the complexity of the localization algorithm. To ensure fairness, all baseline schemes, as well as the proposed APLE-LM and APLE-LM-ACM schemes, use the same search grid $\mathcal{G}$ in subsequent simulations, where the number of $\left\{\chi_{x}, \chi_{y}\right\}$ grid points is fixed at $M_{x} = M_{y} = 4N_{\mathrm{S},x}$.
}
\begin{figure}[t]
    \centering
    \resizebox{0.9\linewidth}{!}{\includegraphics{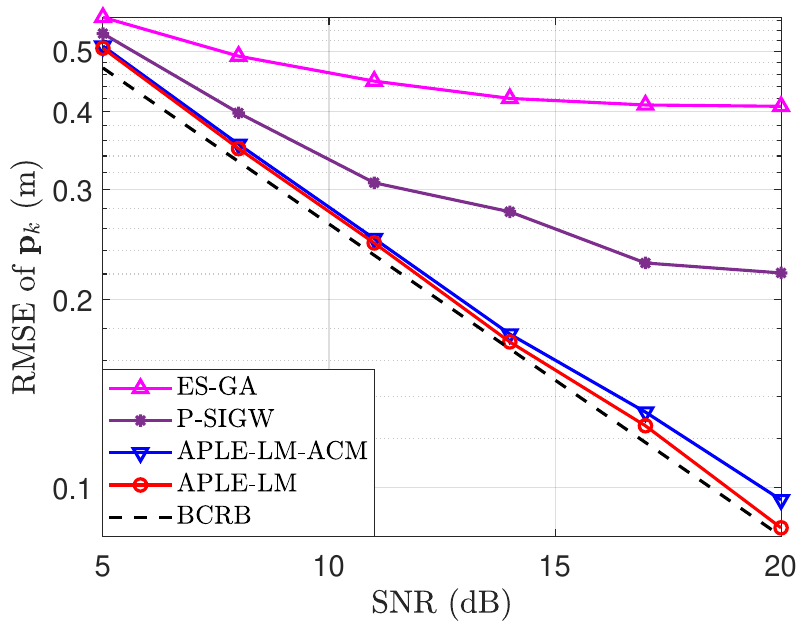}}
    \caption{The UE localization performance v.s. SNR.}
    \label{Fig_positioning}
\end{figure}
\begin{figure}[t]
    \centering
    \resizebox{0.9\linewidth}{!}{\includegraphics{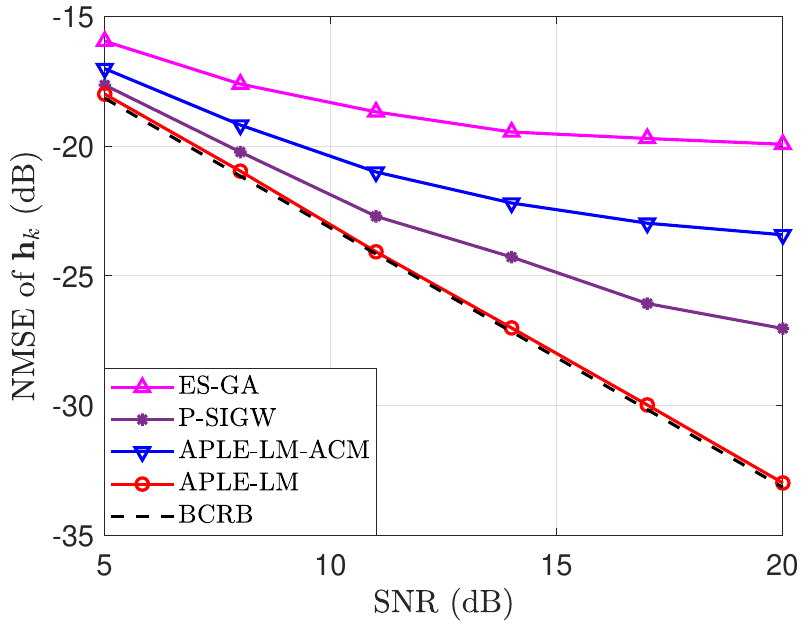}}
    \caption{The channel reconstruction performance v.s. SNR.}
    \label{Fig_channel}
\end{figure}
\subsubsection{Comparation with baseline schemes}
As shown in Fig. \ref{Fig_positioning} and Fig. \ref{Fig_channel}, the proposed algorithm outperforms the baseline schemes in both localization accuracy and channel reconstruction accuracy, approaching the BCRB in the considered SNR range. The baseline schemes, ES-GA and P-SIGW, are sensitive to initialization accuracy, resulting in poor localization performance when a coarse initial grid size is used. In contrast, our algorithm achieves high-precision localization with {a} coarse initial grid size by employing an array partitioning strategy. Additionally, compared to APLE-LM-ACM, the APLE-LM algorithm {employs} a more accurate channel model, resulting in higher precision in both user localization and channel reconstruction. {The performance loss resulting from the approximation in \eqref{eq11} is relatively minor for user localization. However, the performance degradation becomes more pronounced for user position-based channel reconstruction.}

\begin{figure}[t]
    \centering
    \resizebox{0.9\linewidth}{!}{\includegraphics{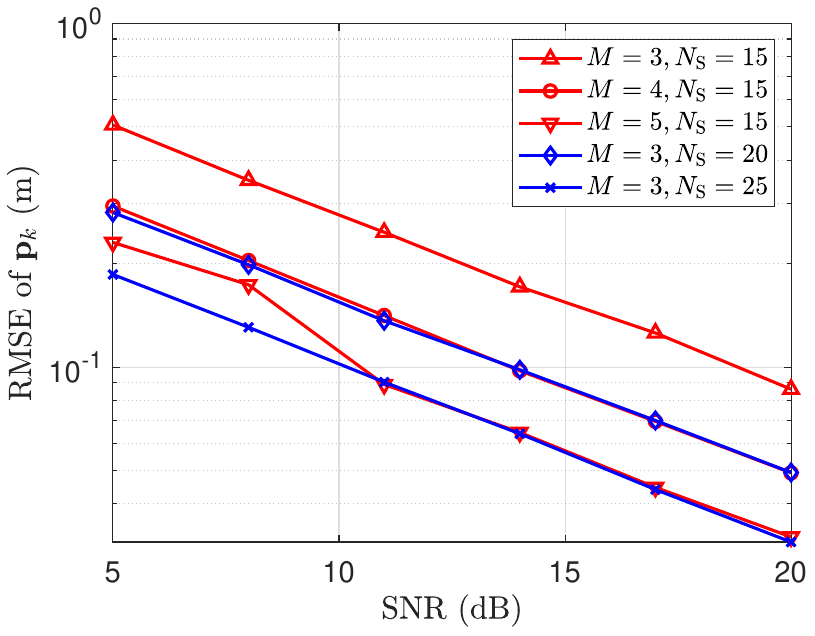}}
    \caption{The UE localization performance v.s. SNR with varying numbers of BS antennas and array partitioning strategy.}
    \label{Fig_arraysize}
\end{figure}
\subsubsection{BS array size}
As shown in Fig. 7, we evaluate the UE localization performance of the proposed APLE-LM algorithm under various antenna array sizes and partitioning strategies. The tested number of array antennas includes $N_{\mathrm{B},x}=45$, 60, and 75. As the number of antennas increases, the RMSE of multi-user positioning decreases, which is due to the higher resolution in both the distance and angle domains provided by larger antenna arrays. Additionally, as shown in Fig. 7, different subarray partitioning strategies yield similar localization performance for the same number of array antennas. For instance, with $N_{\mathrm{B},x}=60$, the $3\times20$ and $4\times15$ partitioning strategies demonstrate comparable performance; similarly, with 75 antennas, the $3\times25$ and $5\times15$ partitioning strategies exhibit nearly identical performance. However, under low SNR, the localization performance of the $5\times15$ partitioning strategy is worse than that of the $3\times25$ strategy. This can be explained by Fig. \ref{fig:4}(b) and Fig. \ref{fig:4}(c), where the messages $\Delta_{\xi\rightarrow\mathbf{p}}(\mathbf{p})$ and $\Delta_{\psi\rightarrow\mathbf{p}}(\mathbf{p})$ exhibit multiple local optima. When the array is partitioned into $5\times15$ subarrays, the initialization errors in the means $\boldsymbol{\mu}_{\xi\rightarrow\mathbf{p}}$ and $\boldsymbol{\mu}_{\psi\rightarrow\mathbf{p}}$ are larger, causing the GD to be trapped in local optima during subsequent message updates, which leads to substantial localization errors.
\begin{figure}[t]
    \centering
    \resizebox{0.9\linewidth}{!}{\includegraphics{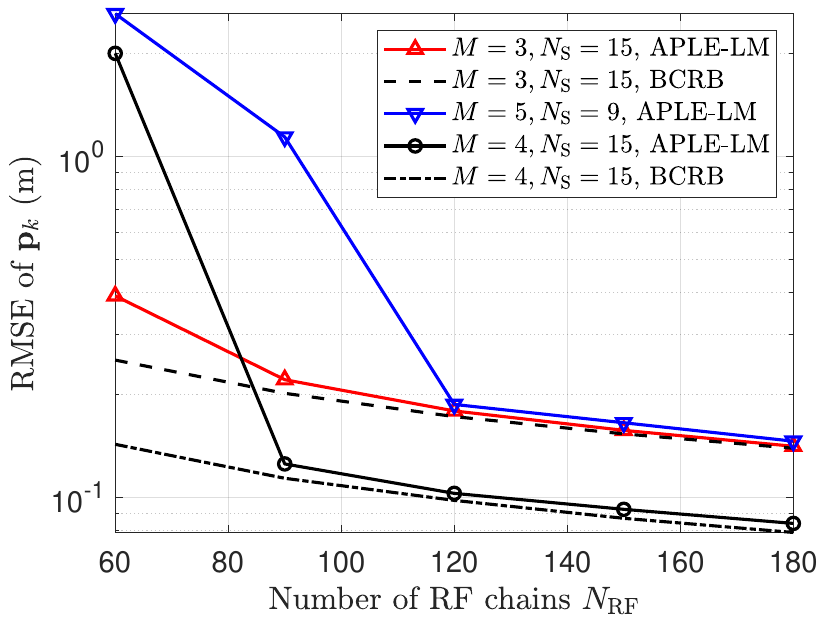}}
    \caption{The UE localization performance v.s. the number of RF chains with varying the array size and partitioning strategies.}
    \label{Fig_RFchian}
\end{figure}
\subsubsection{Number of RF chains}
In Fig. \ref{Fig_RFchian}, we evaluate the user localization performance with different numbers of RF chains at the BS, where the received SNR is fixed at 15 dB. As the number of RF chains increases, providing more BS measurement signals, the localization accuracy consistently improves. However, when the number of RF chains is small {($N_{\mathrm{RF}}<90$)}, there is a significant degradation in localization accuracy for systems configured with $M = 4, N_{\mathrm{S}} = 15$ and $M = 5, N_{\mathrm{S}} = 9$. The cause is similar to the degradation in localization performance observed for the $5\times15$ array partitioning under low SNR. With fewer RF chains, the initialization error in $\boldsymbol{\mu}_{\psi\rightarrow\mathbf{p}}$ is larger, leading $\boldsymbol{\mu}_{\psi\rightarrow\mathbf{p}}$ to be trapped in local optima, resulting in substantial localization errors. {In practice, employing $N_{\mathrm{RF}}=90$ RF chains remains a significant hardware challenge. Under the block fading channel model assumption, we can concatenate the received signals over multiple time slots to increase the number of measurements. Consequently, for a fixed total number of measurements, the required number of RF chains can be further reduced, thereby alleviating hardware complexity.}
\begin{figure}[t]
    \centering
    \resizebox{0.9\linewidth}{!}{\includegraphics{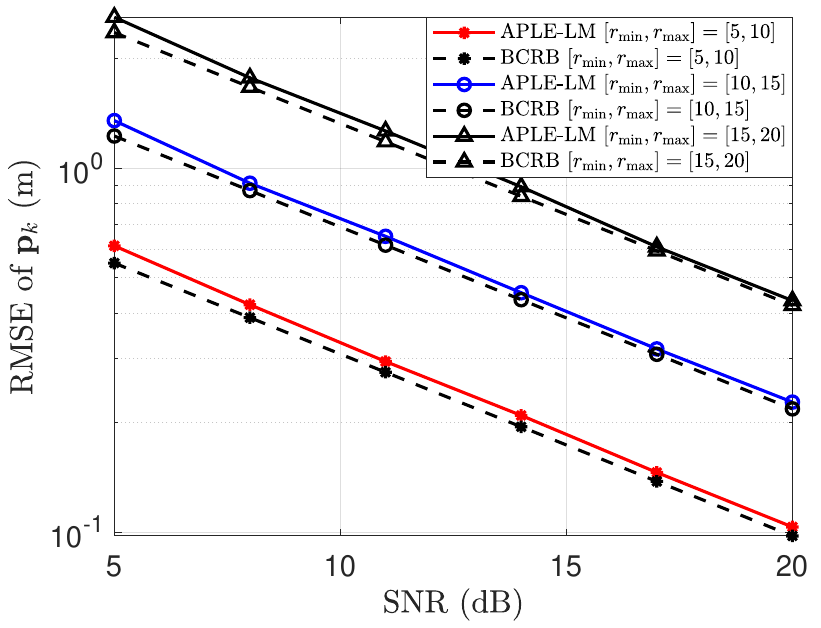}}
    \caption{The UE localization performance v.s. SNR with varying the distance range of UEs.}
    \label{Fig_distance}
\end{figure}
\begin{figure}[t]
    \centering
    \resizebox{0.9\linewidth}{!}{\includegraphics{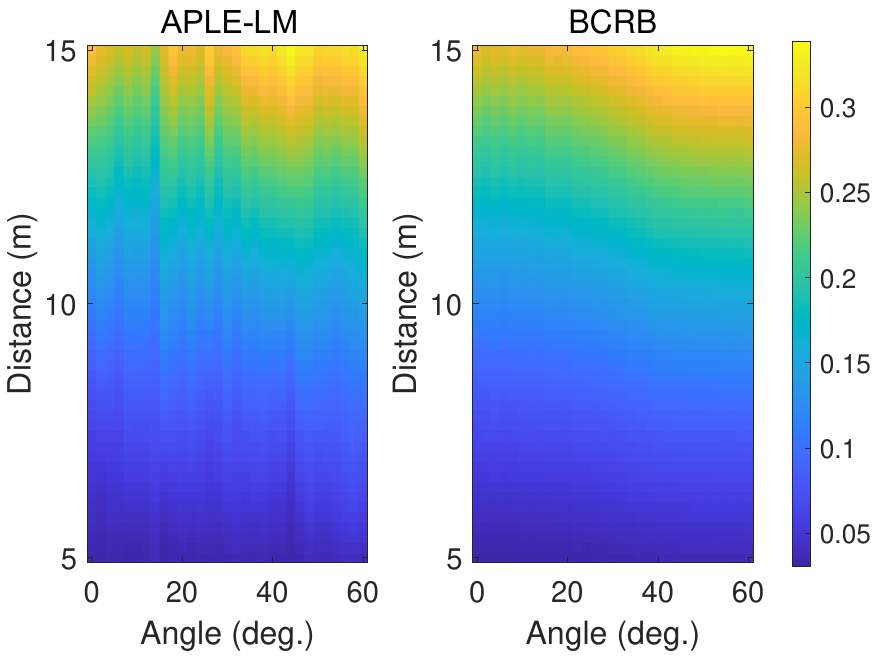}}
    \caption{The UE localization performance v.s. SNR with varying the distance range of UEs.}
    \label{Fig_heat}
\end{figure}
\subsubsection{User locations} We evaluate the proposed APLE-LM algorithm's localization performance for users in different regions. The distance between the BS and the user, $[r_{\mathrm{min}}, r_{\mathrm{max}}]$, is set to $[5,10]$, $[10,15]$, and $[15,20]$ respectively. Fig. \ref{Fig_distance} shows that as the distance between the BS and the user increases, the localization accuracy gradually decreases. In Fig. \ref{Fig_heat}, we consider a scenario where the BS serves a single user. Using polar coordinates to represent the user's position, we evaluate the localization performance of the APLE-LM algorithm and the corresponding CRLB for different user locations. It can be observed that the localization performance degrades as the user's direction deviates from the normal vector of the BS array.
{
\subsubsection{Convergence Behavior} We numerically evaluate the convergence behavior of the proposed algorithm by computing the marginal message $\Delta_{\mathbf{p}}(\mathbf{p})$ in each iteration based on \eqref{eqq52}. The RMSE in \eqref{RMSE} is employed as a metric during the iteration process. As illustrated in Fig. \ref{Fig_conv}, the RMSE of $\mathbf{p}_{k}$ decreases monotonically throughout the iteration process. After a sufficient number of iterations, all curves depicted in Fig. \ref{Fig_conv} eventually converge. Notably, when computing the messages $\Delta_{\mathbf{p}}(\mathbf{p})$ for the first iteration, the localization accuracy is already close to the final accuracy achieved after the algorithm has converged. Therefore, when implementing the APLE-LM algorithm, the number of iterations can be reduced, thereby reducing complexity without causing significant deterioration in localization performance.
\begin{figure}[t]
    \centering
    \resizebox{0.9\linewidth}{!}{\includegraphics{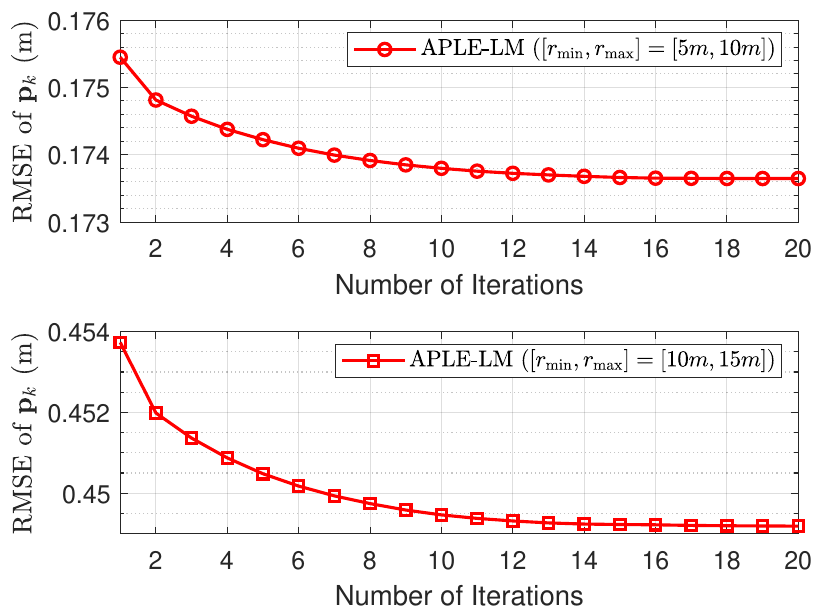}}
    \caption{The convergence behavior of the proposed APLE-LM algorithm. The SNR is fixed at $\SI{14}{dB}$ and two setting of the distance between BS and user are considered, i.e., $[r_{\mathrm{min}},r_{\mathrm{min}}]=[\SI{5}{m},\SI{10}{m}]$ and $[r_{\mathrm{min}},r_{\mathrm{min}}]=[\SI{10}{m},\SI{15}{m}]$.}
    \label{Fig_conv}
\end{figure}
}
\section{Conclusions}\label{S7}
In this paper, we investigated the near-field multiuser localization problem in a multiuser MIMO system, where the BS is equipped with an ELAA. Analog beamforming was employed to reduce the number of RF chains. Due to the large aperture of the ELAA, the user localization problem is highly non-convex and difficult to solve when the number of RF chains is limited. To address this challenge, we modeled the received signals using an array partitioning strategy and constructed a probabilistic model for user location estimation. We proposed the APLE-LM algorithm based on the message-passing principle. We derived the BCRB as the theoretical performance lower bound. Extensive simulations were conducted under various parameter configurations to validate the proposed APLE-LM algorithm. The simulation results demonstrate that the APLE-LM algorithm achieves higher localization accuracy compared to the baseline algorithms and approaches the BCRB at high SNR.
\appendices
\section{Calculation of the Gradient in \eqref{eqq32}}\label{append1}
Based on the chain rule, we obtain
\begin{subequations}
  \begin{align}
    \label{eq655}&\partial\ln\Delta_{\xi\rightarrow\mathbf{p}}\overset{(a)}{=}2\Re\left\{(\partial\boldsymbol{\varXi}_{\xi})^{\mathrm{H}}\boldsymbol{\varUpsilon}_{\xi}^{-1}\boldsymbol{\varXi}_{\xi}\right\}\notag\\
    &\quad\quad-\boldsymbol{\varXi}_{\xi}^{\mathrm{H}}\boldsymbol{\varUpsilon}_{\xi}^{-1}(\partial\boldsymbol{\varUpsilon}_{\xi})\boldsymbol{\varUpsilon}_{\xi}^{-1}\boldsymbol{\varXi}_{\xi}\\
    &\quad=\mathrm{Tr}\{\Re\{\mathbf{Q}^{\mathrm{H}}\partial{\mathbf{B}}\}\}\\
    &\quad\overset{(b)}{=}\mathrm{Tr}\left\{\Re\left\{\sum_{k=1}^{K}\sum_{i=1}^{M^2}\mathbf{q}_{(k-1)M^2+i}^{\mathrm{H}}\mathbf{W}_{i}\partial\mathbf{d}_{k,i}\right\}\right\}
\end{align}
\end{subequations}
where $\mathbf{Q}=2\boldsymbol{C}_{n}^{-1}\left(\boldsymbol{y}-{\mathbf{B}}\boldsymbol{\varUpsilon}_{\xi}^{-1}\boldsymbol{\varXi}_{\xi}\right)\boldsymbol{\varXi}_{\xi}^{\mathrm{H}}\boldsymbol{\varUpsilon}_{\xi,k}^{-1}$, $\mathbf{q}_{j}$ is the $j$-th column of $\mathbf{Q}$,
$(a)$ exploits $\partial (\mathbf{X}^{-1})=-\mathbf{X}^{-1}\partial (\mathbf{X})\mathbf{X}^{-1}$, $(b)$ exploits $\partial{\mathbf{B}}=[\partial \mathbf{B}_{1},...,\partial \mathbf{B}_{K}]$ and $\partial\mathbf{B}_{k}=\left[\mathbf{W}_{1}\partial\mathbf{d}_{k,1},...,\mathbf{W}_{M^2}\partial\mathbf{d}_{k,M^2}\right]$, and $\partial\mathbf{d}_{k,s}=vec(\partial \mathbf{D}_{k}^{(u,v)})$ with the $(i, j)$-th term of $\partial \mathbf{D}_{k}^{(u,v)}$ given by
\begin{align}\label{eq54}
    \partial [\mathbf{D}_{k}^{(u,v)}]_{i,j}=\frac{1}{h_{ref,k}^{(u,v)}}\partial[\boldsymbol{H}_{k}^{(u,v)}]_{i,j}-\frac{[\boldsymbol{H}_{k}^{(u,v)}]_{i,j}}{(h_{ref,k}^{(u,v)})^2}\partial h_{ref,k}^{(u,v)}.
\end{align}
Recall that both $[\boldsymbol{H}_{k}^{(u,v)}]_{i,j}$ and $h_{ref,k}^{(u,v)}$ are elements of channel in \eqref{eq8}, we obtain
\begin{align}\label{eqq65}
    \frac{\partial h_{(i,j),k}}{\partial \mathbf{p}_k} =\alpha_{k}\mathbf{P}_{(i,j),k}\left[\frac{E_{(i,j),k}}{d_{(i,j),k}},\gamma_{(i,j),k}E_{(i,j),k},\frac{\gamma_{(i,j),k}}{d_{(i,j),k}}\right]^{\mathrm{T}},
\end{align}
where $\mathbf{P}_{(i,j),k}=\Big[\frac{\partial\gamma_{(i,j),k}}{\partial \mathbf{p}_k},\frac{\partial\big(\frac{1}{d_{(i,j),k}}\big)}{\partial \mathbf{p}_k},\frac{\partial E_{(i,j),k}}{\partial \mathbf{p}_k}\Big]$, $E_{(i,j),k}\triangleq\exp\left({-\jmath 2\pi \frac{d_{(i,j),k}}{\lambda}}\right)$, and 
\begin{align}
    \frac{\partial\gamma_{(i,j),k}}{\partial \mathbf{p}_k}&=\frac{3}{2}\left(\mathbf{e}_{z}^{\mathrm{T}}\mathbf{e}_{k,i,j}\right)^{\frac{1}{2}}\frac{\mathbf{e}_{z}-\mathbf{e}_{z}^{\mathrm{T}}\mathbf{e}_{k,i,j}\mathbf{e}_{k,i,j}}{d_{(i,j),k}},\\
    \frac{\partial\left(\frac{1}{d_{(i,j),k}}\right)}{\partial \mathbf{p}_k} &= \frac{-\mathbf{e}_{k,i,j}}{d_{(i,j),k}^{2}},\\
    \frac{\partial E_{(i,j),k}}{\partial \mathbf{p}_k}&=\frac{\jmath2\pi}{\lambda}E_{(i,j),k}\mathbf{e}_{k,i,j},\label{eqq68}
\end{align}
with $\mathbf{e}_{k,i,j}=\frac{\mathbf{p}_{k}-\mathbf{q}_{i,j}}{\|\mathbf{p}_{k}-\mathbf{q}_{i,j}\|_{2}}$. Given the polar domain representation in \eqref{eqq32n}, we have
\begin{align}
    \partial \mathbf{p}_k=\mathbf{T}_{k} \left[ \begin{array}{c}
	\partial \chi _{k,x}\\
	\partial \chi _{k,y}\\
	\partial r_k\\
\end{array} \right],
\end{align}
where 
\begin{align}\label{eq62}
    \mathbf{T}_{k}=\left[ \begin{matrix}
	r_k&		0&		\chi _{k,x}\\
	0&		r_k&		\chi _{k,y}\\
	\frac{-r_k\chi _{k,x}}{\sqrt{1-\chi _{k,x}^{2}-\chi _{k,y}^{2}}}&		\frac{-r_k\chi _{k,y}}{\sqrt{1-\chi _{k,x}^{2}-\chi _{k,y}^{2}}}&		\sqrt{1-\chi _{k,x}^{2}-\chi _{k,y}^{2}}\\
\end{matrix} \right].
\end{align}
For $a\in\{\chi _{k,x},\chi _{k,y},r_{k}\}$, we obtain
\begin{align}
    \label{eqq71}\frac{\partial\ln\Delta_{\xi\rightarrow\mathbf{p}}}{\partial a}=\mathrm{Tr}\left\{\Re\left\{\sum_{i=1}^{M^2}\mathbf{q}_{(k-1)M^2+i}^{\mathrm{H}}\mathbf{W}_{i}\frac{\partial\mathbf{d}_{k,i}}{\partial a}\right\}\right\},
\end{align}
where $\frac{\partial\mathbf{d}_{k,j}}{\partial a}$ is calculated by combining \eqref{eq54}-\eqref{eq62}.
\section{Calculation of the Hessian Matrix in \eqref{eq35}}\label{append2}
For arbitrary $a,b\in\{\mathrm{p}_{k,x},\mathrm{p}_{k,y},\mathrm{p}_{k,z}\}$, we have 
\begin{align}
    &\frac{\partial^2\ln\Delta_{\xi\rightarrow\mathbf{p}}}{\partial a\partial b}=\mathrm{Tr}\left\{\Re\left\{\sum_{i=1}^{M^2}\mathbf{q}_{(k-1)M^2+i}^{\mathrm{H}}\mathbf{W}_{i}\frac{\partial^2\mathbf{d}_{k,i}}{\partial a\partial b}\right.\right.\notag\\
    &\quad+\left.\left.\sum_{i=1}^{M^2}\frac{\partial\mathbf{q}_{(k-1)M^2+i}^{\mathrm{H}}}{\partial b}\mathbf{W}_{i}\frac{\partial\mathbf{d}_{k,i}}{\partial a}\right\}\right\},
\end{align}
where $\frac{\partial\mathbf{d}_{k,j}}{\partial a}$ is calculated by combining \eqref{eq54}-\eqref{eqq68}, $\frac{\partial\mathbf{q}_{j}}{\partial b}$ is the $j$-th column of $\frac{\partial \mathbf{Q}}{\partial b}$ calculated in \eqref{eqq73}, and
\begin{figure*}[!ht]
\begin{equation}
    \label{eqq73}\frac{\partial \mathbf{Q}}{\partial b}=-2\boldsymbol{C}_{n}^{-1}\left(\left(\frac{\partial {\mathbf{B}}}{\partial b}\boldsymbol{\varUpsilon}_{\xi}^{-1}\boldsymbol{\varXi}_{\xi}
    + {\mathbf{B}}\frac{\partial\boldsymbol{\varUpsilon}_{\xi}^{-1}}{\partial b}\boldsymbol{\varXi}_{\xi}
    + {\mathbf{B}}\boldsymbol{\varUpsilon}_{\xi}^{-1}\frac{\partial\boldsymbol{\varXi}_{\xi}}{\partial b}
    \right)\boldsymbol{\varXi}_{\xi}^{\mathrm{H}}\boldsymbol{\varUpsilon}_{\xi}^{-1}
    +\left(\boldsymbol{y}-{\mathbf{B}}\boldsymbol{\varUpsilon}_{\xi}^{-1}\boldsymbol{\varXi}_{\xi}\right)
    \left(\left(\frac{\partial\boldsymbol{\varXi}_{\xi}^{\mathrm{H}}}{\partial b}\right)\boldsymbol{\varUpsilon}_{\xi}^{-1}
    +\boldsymbol{\varXi}_{\xi}^{\mathrm{H}}\frac{\partial\boldsymbol{\varUpsilon}_{\xi}^{^{-1}}}{\partial b}\right)\right)
\end{equation}
\hrule
\end{figure*}
\begin{align}
    \frac{\partial\boldsymbol{\varUpsilon}_{\xi}^{^{-1}}}{\partial b} &= -2\boldsymbol{\varUpsilon}_{\xi}^{^{-1}}
    \Re\left\{ {\mathbf{B}}^{\mathrm{H}}
    \boldsymbol{C}_{n}^{-1}
    \frac{\partial {\mathbf{B}}}{\partial b}
    \right\}
    \boldsymbol{\varUpsilon}_{\xi}^{^{-1}},\\
    \frac{\partial\boldsymbol{\varXi}_{\xi}}{\partial b} &= \frac{\partial {\mathbf{B}}^{\mathrm{H}}}{\partial b}\boldsymbol{C}_{n}^{-1}\boldsymbol{y},\\
    \frac{\partial {\mathbf{B}}}{\partial b}&=\left[\mathbf{0},\mathbf{W}_{1}\frac{\partial\mathbf{d}_{k,1}}{\partial b},...,\mathbf{W}_{M^2}\frac{\partial\mathbf{d}_{k,M^2}}{\partial b},\mathbf{0}\right].
\end{align}
$\frac{\partial^2\mathbf{d}_{k,i}}{\partial a\partial b}$ is calculated by
\begin{align}
    &\frac{\partial^2[\mathbf{D}_{k}^{(u,v)}]_{i,j}}{\partial a\partial b} 
    =\frac{1}{h_{ref,k}^{(u,v)}}\frac{\partial^2[\boldsymbol{H}_{k}^{(u,v)}]_{i,j}}{\partial a\partial b}-\frac{[\boldsymbol{H}_{k}^{(u,v)}]_{i,j}}{(h_{ref,k}^{(u,v)})^2}\frac{\partial^2 h_{ref,k}^{(u,v)}}{\partial a\partial b}\notag\\
    &-\frac{1}{\left(h_{ref,k}^{(u,v)}\right)^2}\left(\frac{\partial h_{ref,k}^{(u,v)}}{\partial b}\frac{\partial[\boldsymbol{H}_{k}^{(u,v)}]_{i,j}}{\partial a}
    +\frac{\partial[\boldsymbol{H}_{k}^{(u,v)}]_{i,j}}{\partial b}\frac{\partial h_{ref,k}^{(u,v)}}{\partial a}\right)\notag\\
    &+\frac{2[\boldsymbol{H}_{k}^{(u,v)}]_{i,j}}{(h_{ref,k}^{(u,v)})^3}\frac{\partial h_{ref,k}^{(u,v)}}{\partial b}\frac{\partial h_{ref,k}^{(u,v)}}{\partial a},
\end{align}
where $\frac{\partial h_{ref,k}^{(u,v)}}{\partial a}$ and $\frac{\partial[\boldsymbol{H}_{k}^{(u,v)}]_{i,j}}{\partial a}$ are given in \eqref{eqq65}, $\frac{\partial^2[\boldsymbol{H}_{k}^{(u,v)}]_{i,j}}{\partial a\partial b}$ and $\frac{\partial^2 h_{ref,k}^{(u,v)}}{\partial a\partial b}$ are calculated by
\begin{align}
    &\frac{\partial^2 h_{(i,j),k}}{\partial \mathbf{p}_{k}\partial \mathbf{p}_{k}^{\mathrm{T}}} =\frac{\partial^2\gamma_{(i,j),k}}{\partial \mathbf{p}_{k}\partial \mathbf{p}_{k}^{\mathrm{T}}}\frac{\alpha_{k}E_{(i,j),k}}{d_{(i,j),k}}\notag\\
    &+\alpha_{k}\gamma_{(i,j),k}\frac{\partial^{2} \left(\frac{1}{d_{(i,j),k}}\right)}{{\partial \mathbf{p}_{k}\partial \mathbf{p}_{k}^{\mathrm{T}}}}E_{(i,j),k}
    +\frac{\alpha_{k}\gamma_{(i,j),k}}{d_{(i,j),k}}\frac{\partial^2 E_{(i,j),k}}{{\partial \mathbf{p}_{k}\partial \mathbf{p}_{k}^{\mathrm{T}}}}\notag\\
    &+ \alpha_{k}\mathbf{P}_{(i,j),k}\mathbf{M}_{(i,j),k}\mathbf{P}_{(i,j),k}^{\mathrm{T}},
\end{align}
with $\frac{\partial^2\gamma_{(i,j),k}}{\partial \mathbf{p}_{k}\partial \mathbf{p}_{k}^{\mathrm{T}}}$ given in \eqref{eq79} and
\begin{figure*}[!ht]
\begin{equation}
\label{eq79}
    \frac{\partial^2\gamma_{(i,j),k}}{\partial \mathbf{p}_{k}\partial \mathbf{p}_{k}^{\mathrm{T}}}=\frac{3\left(\mathbf{I}-\mathbf{e}_{k,i,j}\mathbf{e}_{k,i,j}^\mathrm{T}\right)\mathbf{e}_{z}\mathbf{e}_{z}^{\mathrm{T}}\left(\mathbf{I}-\mathbf{e}_{k,i,j}\mathbf{e}_{k,i,j}^\mathrm{T}\right)}{4\left(\mathbf{e}_{z}^{\mathrm{T}}\mathbf{e}_{k,i,j}\right)^{\frac{1}{2}}d_{(i,j),k}^2}
    +\frac{3\left(\mathbf{e}_{z}^{\mathrm{T}}\mathbf{e}_{k,i,j}\right)^{\frac{1}{2}}}{2d_{(i,j),k}^2}
    \left(\mathbf{e}_{z}^{\mathrm{T}}\mathbf{e}_{k,i,j}(3\mathbf{e}_{k,i,j}\mathbf{e}_{k,i,j}^{\mathrm{T}}-\mathbf{I})-\mathbf{e}_{k,i,j}\mathbf{e}_{z}^{\mathrm{T}}-\mathbf{e}_{z}^{\mathrm{T}}\mathbf{e}_{k,i,j}\right).
\end{equation}
\hrule
\end{figure*}
\begin{align}
    &\frac{\partial^{2} \left(\frac{1}{d_{(i,j),k}}\right)}{{\partial \mathbf{p}_{k}\partial \mathbf{p}_{k}^{\mathrm{T}}}}=\frac{3\mathbf{e}_{k,i,j}\mathbf{e}_{k,i,j}^{\mathrm{T}}-\mathbf{I}}{d_{(i,j),k}^{3}},\\
    &\frac{\partial^2 E_{(i,j),k}}{{\partial \mathbf{p}_{k}\partial \mathbf{p}_{k}^{\mathrm{T}}}}=\frac{\jmath2\pi}{\lambda}E_{(i,j),k}\left(\frac{\jmath2\pi}{\lambda}\mathbf{e}_{k,i,j}\mathbf{e}_{k,i,j}^{\mathrm{T}}
    +\frac{\mathbf{I}-\mathbf{e}_{k,i,j}\mathbf{e}_{k,i,j}^{\mathrm{T}}}{d_{(i,j),k}}\right),\\
    &\mathbf{M}_{(i,j),k}=\left[ \begin{matrix}
	0&		E_{(i,j),k}&		\frac{1}{d_{(i,j),k}}\\
	E_{(i,j),k}&		0&		\gamma_{(i,j),k}\\
	\frac{1}{d_{(i,j),k}}&		\gamma_{(i,j),k}&		0\\
\end{matrix} \right].
\end{align}
When $a$ and $b$ are from different users, i.e., $a\in\{\mathrm{p}_{k_1,x},\mathrm{p}_{k_1,y},\mathrm{p}_{k_1,z}\}$ and $b\in\{\mathrm{p}_{k_2,x},\mathrm{p}_{k_2,y},\mathrm{p}_{k_2,z}\}$, the Hessian matrix is given by
\begin{align}
    \frac{\partial^2\ln\Delta_{\xi\rightarrow\mathbf{p}}}{\partial a\partial b}=\mathrm{Tr}\left\{\Re\left\{\sum_{i=1}^{M^2}\frac{\partial\mathbf{q}_{(k_2-1)M^2+i}^{\mathrm{H}}}{\partial b}\mathbf{W}_{i}\frac{\partial\mathbf{d}_{k_1,i}}{\partial a}\right\}\right\}.
\end{align}
\bibliographystyle{IEEEtran}
\bibliography{mybib}
\end{document}